\documentclass[journal]{IEEEtran}
\usepackage[T1]{fontenc}
\usepackage[pdftex]{graphicx}
\usepackage{algorithm,algorithmic}
\usepackage{amsmath}
\usepackage{cite}
\usepackage{multirow}
\usepackage{mathtools, cuted}
\usepackage{booktabs}
\usepackage{xcolor}

\usepackage[utf8]{inputenc}
\usepackage[english]{babel}

\usepackage{cleveref}
\usepackage{ amssymb }
\usepackage{subcaption}
\usepackage{float}
\usepackage{todonotes}
\usepackage{verbatim} 
\usepackage{tabularx}
\usepackage{enumitem}

\newcommand{\bs}[1]{{\boldsymbol #1}}
\newcommand{\Exp}[1]{\mathbb{E}\left\{#1\right\}}
\newcommand{\hs}[1]{\hspace{-#1 ex}}
\begin{document}

\title{Channel Estimation for Full-duplex Multi-tag Ambient Backscatter Communication Systems with I/Q Imbalance}

\author{Saeed Abdallah$^*$,~\IEEEmembership{~Member,~IEEE}, Mahmoud A. Albreem,~\IEEEmembership{Senior Member,~IEEE}, Bassel Al Homssi,~\IEEEmembership{~Senior Member,~IEEE} Mohamed Saad,~\IEEEmembership{~Senior Member,~IEEE}, Abdulmalik Alwarafy,~\IEEEmembership{~Member,~IEEE}

\thanks{\noindent Saeed Abdallah and Mahmoud A. Albreem are with the Department of Electrical Engineering, College of Engineering, University of Sharjah, Sharjah, UAE, e-mail: \texttt{sabdallah,malbreem@sharjah.ac.ae}.}
\thanks{\noindent Bassel Al Homssi and Mohamed Saad are with the Department of Computer Engineering, College of Computing and Informatics, University of Sharjah, Sharjah, UAE, e-mail: \texttt{bhomssi,msaad@sharjah.ac.ae}.}
\thanks{\noindent Abdulmalik Alwarafy is with the Department of Computer and Network Engineering, College of Information Technology, United Arab Emirates University, Al Ain, UAE, e-mail: \texttt{aalwarafy@uaeu.ac.ae}.}
\thanks{$^*$ Corresponding author.}

\thanks{This research was supported in part by the Distributed and Networked Systems (DNS) Research Group, Smart Automation and communication Technologies (SACT) Research Center, University of Sharjah, Operating Grant no. 150410 and in part by Competitive Grant no. 23020403289.}
}

\maketitle

\begin{abstract}

Ambient backscatter communication (AmBC) has emerged as a highly attractive paradigm for energy-efficient communication. Full-duplex multi-tag AmBC systems provide the scalability and efficient spectrum utilization essential for next generation Internet-of-Things (IoT) networks. However, the presence of multiple tags, self-interference and hardware impairments such as inphase/quadrature (I/Q) imbalance, makes accurate channel estimation indispensable for efficient interference management. The large number of channel parameters and the presence of mirror images of each signal component necessitate careful design of the channel estimation phase to prevent performance degradation. In this work, we propose a novel three-stage training protocol and pilot-based estimation scheme that ensure signal orthogonality and successfully avoid error floors. We also propose two semi-blind estimators, one based on decision-directed (DD) criterion and the other on the expectation conditional maximization (ECM) framework. By exploiting both pilots and data symbols, these two estimators achieve higher estimation accuracy than pilot-based estimation, at the cost of additional complexity. Cramer-Rao bounds (CRBs) for both types of estimation are also derived. The pilot-based estimator and the ECM estimator approach their respective CRBs, while the DD estimator performs mid-way between them. The three proposed solutions support different use cases by offering distinct tradeoffs between performance and computational complexity.

\end{abstract}


\IEEEpeerreviewmaketitle

\section{Introduction}

The rising demands of the Internet-of-Things (IoT) have especially highlighted the need for more robust wireless communication~\cite{9145564}, to handle the pervasiveness of connected devices and the massive volumes of data transmitted in real time. Future networks are expected to provide ultra-low latency, very high throughput, and continuity. Next-generation wireless networks, such as 5G and beyond, are being developed to fulfill these varied needs. They implement features including massive MIMO, mmWave communications, and powerful coding to enhance data throughput and extend network coverage~\cite{9395074}. While these techniques improve the performance of wireless systems, energy efficiency is usually lagging behind~\cite{9632691}. Most IoT deployments involve battery-operated devices that have a small energy source. Consequently, future communication paradigms must be designed to meet the performance-power trade off in order to enable next-generation networks to operate efficiently in an energy-constrained manner.

Ambient backscatter communication (AmBC) is a novel paradigm with the potential to revolutionize wireless communication by enabling ultra-low-power connected pervasive IoT devices~\cite{8368232}. AmBC enables devices to transmit data by modulating and reflecting ambient RF signals that are always present in the environment~\cite{8368232}. A tag uses these signals—Wi-Fi, cellular, and satellite signals—to create its own signal by altering the reflectivity of the device’s antenna according to the data to be sent, where nearby receivers can then pick up these variations~\cite{8103807,7551180}. AmBC systems can be operated with low cost and very low power consumption, making them ideal for long-lifetime IoT applications.

While non-coherent operation has been considered in traditional AmBC, both cooperative and full-duplex AmBC systems require accurate channel state information (CSI). In the former, the CSI is needed to recover the information of both the RF source and the tag from the superposed signal. In the latter, channel estimation is additionally needed to handle the residual self-interference (RSI). The need for accurate CSI becomes even more evident in the multi-tag case where it is essential to separate the overlapping backscattered signals.

\subsection{Related Work}

Researchers have sought to provide efficient solutions to the AmBC channel estimation problem. For traditional AmBC systems, blind channel estimation was proposed in~\cite{8320359} using the expectation maximization (EM) algorithm. In cooperative AmBC systems\cite{8274950}, the RF source can collaborate with the tags and the reader, enabling the transmission of pilot sequences. Hence, in~\cite{8927865} pilot sequences and an iterative estimator were proposed for AmBC systems where the reader is equipped with a large intelligent surface. EM-based semi-blind estimation for the cooperative scenario was proposed in~\cite{8689198}. The case of a massive-antenna equipped reader was studied in~\cite{8746230}. Using pilot transmission from the source, a two-stage algorithm was proposed to jointly estimate the channels and direction-of-arrival (DoA). Pilot-based channel estimation was cast as a denoising problem in~\cite{9222226}, and convolutional neural network (CNN)-based deep residual learning was employed, achieving close performance to the minimum mean squared error (MMSE) method. Pilot-based channel estimation was also cast as a denoising problem in~\cite{10555303}, where conditional generative adversarial networks were employed to leverage the spatial and temporal features of the pilot signals, leading to improved performance, especially at low SNR. Channel estimation for full-duplex AmBC systems with multi-antenna readers was studied in~\cite{10057435}. Both the AP and the legacy user (LU) were assumed to operate in full-duplex mode, which resulted in the backscattering of both their signals. In addition to a pilot-based least-squares (LS) estimator, two semi-blind estimators were explored, one based on the decision-directed (DD) strategy and the other on the EM framework. The semi-blind methods yielded superior accuracy and required lower training overhead, but had higher computational complexity.

The above methods did not take into consideration the impact of hardware impairments such as inphase/quadrature (I/Q) imbalance and carrier frequency offset (CFO). Ambient backscatter readers and tags are typically built around low-cost direct-conversion (zero-IF) radios to minimize size, cost, and power. In such radios, the RF signal is mixed directly to baseband using quadrature oscillators, making them particularly vulnerable to I/Q imbalance. Specifically, any gain or phase mismatch between the in-phase and quadrature branches directly creates image-frequency interference at baseband. This impairment is intrinsic to the front-end architecture and cannot be eliminated by filtering. I/Q imbalance is particularly noteworthy for introducing mirror images of signals, which further exacerbate interference and increase the number of parameters to be estimated~\cite{10480412}. This issue was addressed in~\cite{9415632}, where the channel, I/Q imbalance and CFO were jointly estimated using both pilot-based and EM-based techniques. Joint estimation of the channel and I/Q imbalance was also studied for OFDM-based full-duplex AmBC systems in~\cite{10480412}, where appropriate training was designed and both a pilot-based LS method and a DD semi-blind method were proposed. The former had lower complexity, while the latter yielded higher estimation accuracy. Joint estimation of the CFO and the sampling frequency offset (SFO) for AmBC systems was considered in~\cite{10726883}, where appropriate pilot sequences were designed to optimize the joint estimation. 

The above works mostly assume a single tag. The multi-tag scenario is more challenging since it requires the estimation of multiple interfering cascaded channels from the superposition of backscattered signals. Although it is possible to apply pilot-based single-tag algorithms by activating only a single tag at a time, this is highly inefficient and imposes an impractical training overhead~\cite{10271380}. Consequently, the authors of~\cite{10271380} proposed several pilot designs to simultaneously estimate the channel in the multi-tag scenario, showing that the joint scheme outperforms single-tag techniques.

\subsection{Contribution and Organization}

 Existing works on multi-tag channel estimation have mainly considered pilot-based estimation\footnote{The authors published a preliminary study on semi-blind channel estimation for multi-tag AmBC systems in the conference paper~\cite{10812591}, however the work therein considered a half-duplex system with perfect I/Q balance.}~\cite{10271380}. Semi-blind estimation is capable of achieving higher accuracy, lower pilot overhead and superior spectrum utilization~\cite{8966610}. Furthermore, the work in~\cite{10271380} does not consider the impact of I/Q imbalance, which can introduce mirror images of the direct component and backscattered component, resulting in significant performance degradation and requiring accurate compensation. Moreover, the work in~\cite{10271380} considers half-duplex AmBC. Full-duplex AmBC would add extra complexity due to the presence of RSI and the backscattering of both the AP and LU signals.

In this work, we address the problem of channel estimation in a multi-tag full-duplex AmBC system that is affected by both transmit (TX) and receive (RX) I/Q imbalance. In our work (like~\cite{10271380}), we assume that the tags transmit PSK symbols, whereby they can modify the phase of the backscattered signal. The use of complex (phase-shifted) tag symbols in the presence of TX and RX I/Q imbalances completely transforms the received signal. In particular, the received signal corresponding to each backscattered component involves a total of four components, in contrast to the two components considered in~\cite{10480412} and~\cite{10485514}. This necessitates the estimation of four channel components for each tag. Additionally, the fact that both the LU and AP operate in full-duplex mode means that the tags backscatter the signals from both the LU and the AP, as in~\cite{10057435} and~\cite{10480412}. Therefore, for each tag a total of eight channel components need to be estimated. The resulting estimation problem is highly challenging and cannot be solved using simple combinations of existing methods. Designing appropriate pilot sequences is necessary to guarantee orthogonality among the signal components during the training and avoid potential error floors.

In addressing the above challenges, we consider both pilot-based and semi-blind estimation strategies. We first develop a training protocol consisting of three phases, and select the training sequences in each phase to guarantee orthogonality among all signal components. We also develop a pilot-based estimation algorithm that goes hand-in-hand with the above protocol. As a benchmark, the Cramer-Rao bound (CRB) for pilot-based estimation is obtained. Importantly, the performance of the proposed algorithm coincides with the CRB.

In addition, we develop two semi-blind estimation methods, which offer different performance/complexity tradeoffs. The first semi-blind estimator is based on the DD estimation strategy, whereby the pilot-based estimator is used to detect a sequence of transmitted data symbols, and then another estimation stage is applied using both the original pilots and the detected data symbols. Alternating maximization is used to facilitate the estimation of all parameters, and the resulting scheme is referred to as the AMDD estimator. While it yields improved accuracy over the pilot-based estimation, it can suffer from error propagation, which limits the performance improvement at low SNR. As a remedy, another semi-blind algorithm is proposed based on the expectation conditional maximization (ECM) approach. This iterative method uses the posterior probabilities of the data symbols, allowing it to achieve superior performance. The semi-blind modified CRB (MCRB) is also obtained as a benchmark for semi-blind estimators. The ECM performs closest to the MCRB, almost coinciding with it at high SNR.

The contributions can be summarized as follows: 
\begin{itemize}
\item A novel training protocol is developed for full-duplex multi-tag AmBC systems with I/Q imbalance and a multi-antenna reader, guaranteeing orthogonality among all signal components during the training.
\item A pilot-based estimator is developed that goes hand-in-hand with the above protocol and enables the estimation of all channel parameters without error floors.
\item A DD-based semi-blind estimator is proposed, achieving higher accuracy than the pilot-based estimator. 
\item Another semi-blind estimator is developed based on the ECM criterion. This estimator yields the highest accuracy. 
\item As theoretical benchmarks, both the pilot-based and semi-blind CRBs are obtained. The pilot-based estimator coincides with the pilot-based CRB, while the ECM converges to the semi-blind CRB at high SNR.
\item Extensive simulation results are used to illustrate the performances of proposed algorithms, and investigate various tradeoffs.
\end{itemize}

The rest of this paper is organized as follows. Section~\ref{system_model} presents our system model. The training protocol and pilot-based estimation are presented in Section~\ref{pilot_ML}. The semi-blind AMDD estimator is developed in Section~\ref{proposed_AMDD}. The ECM estimator is developed in Section~\ref{proposed_ECM_estimators}. The computational complexity of the proposed estimators is analyzed in Section~\ref{computational_complexity}. The pilot-based and semi-blind CRBs are obtained in Sections~\ref{Pilot_CRB} and~\ref{SB_CRB}, respectively. Simulation results are provided in Section~\ref{simulations}. Finally our conclusions are in Section~\ref{conclusions}.

\textit{Notation:} For a column vector $\bs{a}$, $\Vert\bs{a}\Vert$ denotes the two-norm, while $a[i]$ denotes the $i$th element. For a matrix $\bs{A}$, the transpose and Hermitian are denoted by $\bs{A}^{T}$, $\bs{A}^{H}$,  respectively. $\bs{A}^{\dagger}$ denotes the Moore-Penrose pseudo-inverse of $\bs{A}$.  For a complex number $c$, $\Re\{c\}$, $\Im\{c\}$ and $c^{*}$ represent the real part, imaginary part and conjugate, respectively. $\bs{I}_N$ represents the $N\times N$ identity matrix. The notation $\mathcal{CCN}(0,\sigma^2\bs{I}_N)$ describes a circularly symmetric complex Gaussian random vector with mean zero and covariance matrix $\sigma^2\bs{I}_N$. For a vector $\bs{a}$, $\mbox{Diag}(\bs{a})$ is the diagonal matrix whose diagonal elements are the elements of $\bs{a}$. $\bs{A}\otimes\bs{B}$ represents the Kronecker product between matrices $\bs{A}$ and $\bs{B}$. The operation $\mbox{vec}(\bs{B})$ transforms a matrix $\bs{B}$ into a column vector, by sequentially concatenating its columns. $\bs{x}\odot\bs{y}$ denotes the element-wise multiplication of the vectors $\bs{x}$ and $\bs{y}$.

\section{System Model}
\label{system_model}

\begin{figure}[t]
    \centering
    \captionsetup{type=figure, justification=centering}
    \includegraphics[width=0.8\linewidth]{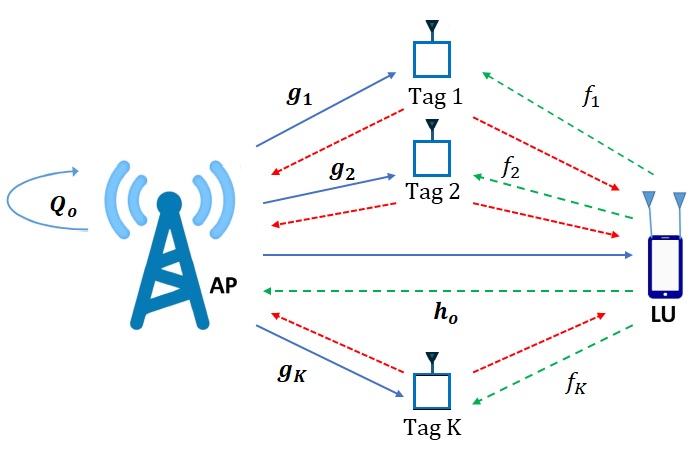}
    \caption{Full-duplex multi-tag AmBC system under consideration (excluding the impact of I/Q imbalance).}
    \label{system_diagram}
\end{figure}

We consider the multi-tag AmBC system in Fig.~\ref{system_diagram}, where $K$ tags communicate with an $M$-antenna AP, while the AP communicates with a LU. As in~\cite{10057435} and~\cite{10480412}, we assume that both the AP and the LU operate in full-duplex mode. Hence, the tags reflect the signals of both the AP and the LU. Furthermore, as in~\cite{10271380}, we assume that the tags use PSK, whereby they vary their antenna impedances to introduce desired phase shifts in the backscattered signals.

We assume that the system is affected by both transmit (TX) and receive (RX) I/Q imbalances. Suppose that $x$ is the baseband signal to be transmitted under perfect I/Q balance. In the presence of TX I/Q imbalance, the actual transmitted signal is given by
$\tilde{x}=G_1x+G_2x^{*}$,
where $G_1=(1+g_Te^{j\phi_T})/2$ and $G_2=(1-g_Te^{-j\phi_T})/2$, where $g_T$ and $\phi_T$ represent the amplitude and phase imbalances, respectively. 

Similarly, on the receiver side, suppose that $r$ is the baseband (downconverted) RX signal under perfect RX I/Q balance. Then the actual received signal is given by $\tilde{r}=K_1r+K_2r^{*}$,
where $K_1=(1+g_Re^{j\phi_R})/2$ and $K_2=(1-g_Re^{-j\phi_R})/2$, where $g_R$ and $\phi_R$ represent the amplitude and phase imbalances, respectively. 

The distance between the LU and the AP is denoted as $d_R$, whereas the distance between the LU and the $k$th tag is denoted as $\bar{d}_k$, and that between the $k$th tag and the AP is $d_k$. The pathloss corresponding to a distance $d$ is modeled as
\begin{equation}
\label{pathloss}
L(d)=\left(\frac{\lambda}{4\pi d_o}\right)^2(d_o/d)^\gamma,
\end{equation}
where $\lambda$ is the wavelength of the signal, $d_o$ is the reference distance and $\gamma$ is the pathloss exponent. The small-scale fading is modeled using Nakagami-$m$ fading, a versatile model that covers both LOS and NLOS fading scenarios.

During the $n$th time slot, the LU transmits the signal $s[n]$, while the $k$th tag uses backscattering to transmit the signal $t_k[n]$, and the AP transmits the $M\times1$ signal $\bs{r}[n]$. The corresponding $M\times1$ received signal vector at the AP is 
\begin{equation}
\label{RXwithIQ}
\tilde{\bs{y}}[n]=K_1\bs{y}[n]+K_2\bs{y}^{*}[n],
\end{equation}
where $\bs{y}[n]$ is the received signal under perfect RX I/Q balance, given by
\begin{equation}
\label{rx_sig1}
\begin{split}
    \bs{y}[n] =& \sqrt{L(d_R)}(\bs{a}\tilde{s}[n] + \sum\limits_{k=1}^{K}\sqrt{L(d_k)L(\bar{d}_k)}\eta_kf_k\bs{g}_k \tilde{s}[n]t_k[n] \\
    &+\bs{Q}_o\tilde{\bs{r}}[n] + \sum\limits_{k=1}^{K}L(d_k)\eta_k\bs{g}_k\bs{g}_k^T t_k[n]\tilde{\bs{r}}[n] + \bs{w}[n], 
    \end{split}
\end{equation}
where $\tilde{s}[n]=G_1s[n]+G_2s^{*}[n]$, $\tilde{\bs{r}}[n]=G_1\bs{r}[n]+G_2\bs{r}^{*}[n]$ and $0<\eta_k<1$ denotes the reflection coefficient of the $k$th tag. Moreover, the $M\times1$ vector $\bs{a}$ represents the small-scale fading between the LU and the AP, while $f_k$ represents the small-scale fading  between the LU and the $k$th tag, and the $M\times1$ vector $\bs{g}_k$ represents the small-scale fading between the $k$th tag and the AP. Furthermore, the $M\times M$ matrix $\bs{Q}_o$ denotes the RSI channel at the AP, while $\bs{w}[n]$ represents the additive white Gaussian noise (AWGN), which is $\mathcal{CCN}(\bs{0},\sigma^2\bs{I}_M)$.

Letting $\bs{h}_o\triangleq \sqrt{L(d_R)}\bs{a}$, $\bs{v}_k\triangleq \sqrt{L(d_k)L(\bar{d}_k)}\eta_kf_k\bs{g}_k$ and $\bs{U}_k\triangleq L(d_k)\eta_k \bs{g}_k\bs{g}_k^T$, the signal $\bs{y}[n]$ can be expressed as
\begin{equation}
\label{rx_sig2}
\begin{split}
    \bs{y}[n] =& \bs{h}_o\tilde{s}[n] + \sum\limits_{k=1}^{K}\bs{v}_k \tilde{s}[n]t_k[n] +\bs{Q}_o\tilde{\bs{r}}[n] \\&+ \sum\limits_{k=1}^{K}t_k[n]\bs{U}_k\tilde{\bs{r}}[n] + \bs{w}[n]. 
    \end{split}
\end{equation}

To simplify our notation, we define the following effective channel coefficients that simultaneously account for the impact of the fading and I/Q imbalance:
\begin{equation*}
\bar{\bs{h}}\triangleq K_1G_1\bs{h}_o+K_2G_2^{*}\bs{h}_o^{*},\ \ 
\check{\bs{h}}\triangleq K_1G_2\bs{h}_o+K_2G_1^{*}\bs{h}_o^{*},
\end{equation*}
\begin{equation*}
\bar{\bs{Q}}\triangleq K_1G_1\bs{Q}_o+K_2G_2^{*}\bs{Q}_o^{*},\ \
\check{\bs{Q}}\triangleq K_1G_2\bs{Q}_o+K_2G_1^{*}\bs{Q}_o^{*},
\end{equation*}
\begin{equation*}
\bar{\bs{v}}_k\triangleq K_1G_1\bs{v}_k,\ \
\check{\bs{v}}_k\triangleq K_2G_1^{*}\bs{v}_k^{*},\ \
\dot{\bs{v}}_k\triangleq K_2G_2^{*}\bs{v}_k^{*},
\end{equation*}
\begin{equation*}
\ddot{\bs{v}}_k\triangleq K_1G_2\bs{v}_k,\ \
\bar{\bs{U}}_k\triangleq K_1G_1\bs{U}_k,\ \
\check{\bs{U}}_k\triangleq K_2G_1^{*}\bs{U}_k^{*},
\end{equation*}
\begin{equation*}
\dot{\bs{U}}_k\triangleq K_2G_2^{*}\bs{U}_k^{*},\ \
\ddot{\bs{U}}_k\triangleq K_1G_2\bs{U}_k.
\end{equation*}

We may now express $\tilde{\bs{y}}[n]$ in~\eqref{RXwithIQ} as
\begin{equation}
\label{overall_sig}
\begin{split}
&\tilde{\bs{y}}[n]=\bar{\bs{h}}s[n]+\check{\bs{h}}s[n]^{*}+\bar{\bs{Q}}\bs{r}[n]+\check{\bs{Q}}\bs{r}[n]^{*}\\
&+\sum\limits_{k=1}^{K}\Big(\bar{\bs{v}}_ks[n]t_k[n]+\dot{\bs{v}}_ks[n]t_k^{*}[n]
+\ddot{\bs{v}}_ks[n]^{*}t_k[n]\\&+\check{\bs{v}}_ks[n]^{*}t_k[n]^{*}\Big) 
+\sum\limits_{k=1}^{K}\Big(\bar{\bs{U}}_kr[n]t_k[n]+\dot{\bs{U}}_kr[n]t_k^{*}[n]\\&
+\ddot{\bs{U}}_kr[n]^{*}t_k[n]+\check{\bs{U}}_kr[n]^{*}t_k[n]^{*}\Big)+\tilde{\bs{w}}[n],
\end{split}
\end{equation}
where $\tilde{\bs{w}}[n]\triangleq K_1\bs{w}[n]+K_2\bs{w}[n]^{*}$. We note that $\tilde{\bs{w}}[n]$ is $\mathcal{CCN}(\bs{0},\tilde{\sigma}^2\bs{I}_M)$, where $\tilde{\sigma}^2\triangleq(|K_1|^2+|K_2|^2)\sigma^2$. Obviously, estimating the channel coefficients in~\eqref{overall_sig} is a challenging task that requires careful pilot design to handle the superposition between the LU signal, its mirror image, the RSI signal, its mirror image, as well as the backscattered signals and their mirror images. The fact that we consider PSK transmission at the tags as well as both TX and RX I/Q imbalances gives rise to 3 additional mirror images of each backscattered signal, further complicating the estimation problem.

 For the RSI channel $\mathbf{Q}_o$, self-interference typically comprises both line-of-sight (LOS) and non-line-of-sight (NLOS) components. Because the LOS part varies only slowly and can be largely suppressed by RF-cancellation methods, we treat the residual self-interference as arising mainly from scattering. Consequently, $\mathbf{Q}_o$ is modeled as a Rayleigh-fading channel whose entries follow $\mathcal{CCN}(0,\sigma_i^2)$ \cite{li2018optimal}.

In our work we assume that the LU, the AP and the tags are synchronized. Synchronization techniques for AmBC systems have already been investigated in works such as~\cite{zhang2016hitchhike} and~\cite{dunna2021syncscatter}. Specifically, in~\cite{dunna2021syncscatter} an integrated circuit is presented to achieve synchronized backscatter communication with ambient Wi-Fi signals, supporting multiple tags.

Our work assumes that the tags perform energy harvesting (EH) and data transmission concurrently~\cite{10555303}. Hence, each tag splits the received RF signal power into two parts based on a power splitting ratio $\lambda$. A linear model is commonly assumed in EH circuits, whereby the harvested power is a linear function of the input power, i.e., $P_h=\delta\lambda P_r$, where $0<\delta<1$ is the power conversion efficiency. The energy harvesting process, however, does not have a major bearing on the channel estimation algorithms, as pointed out in~\cite{10555303}.

\section{Pilot-Design and Pilot-based Channel Estimation}
\label{pilot_ML}

In this section, we propose a pilot-based estimation scheme and an appropriate design of pilot sequences that together enable the estimation of all the required channel coefficients. It is very difficult to estimate the channel coefficients simultaneously in one shot, due to the interference and coupling among the different components. The transmitted LU signal appears both individually, as well multiplied by the signals of all the tags, and their mirror images. The same applies to the transmitted AP signal. Orthogonality also needs to hold across the signals of the multiple tags. To effectively address this problem, we divide our proposed scheme into three stages. 

\subsubsection{Pilot Training Phase 1}

The first training phase aims to estimate the direct and self-interference channels. To avoid interference from the backscattered signals, all the tags are assumed to be in the absorbing state. We let $\bs{s}_1\triangleq\big[s_1[1],\hdots ,s_1[N_1]\big]^{T}$ be the sequence of training symbols transmitted by the LU. We also let $\bs{r}_m\triangleq\big[r_m[1],\hdots r_m[N_1]\big]^T$ be the sequence of training symbols transmitted by the $m$th antenna of the AP during the same itnerval. We also define the $N_1\times M$ matrix $\bs{R}\triangleq \big[\bs{r}_1,\hdots,\bs{r}_M\big]$. The corresponding $M\times N_1$ received signal at the AP is expressed as
\begin{equation}
\label{EqY0}
\bs{Y}_1=\bar{\bs{h}}\bs{s}_1^{T}+\check{\bs{h}}\bs{s}_1^{H}+\bar{\bs{Q}}\bs{R}^{T}+\check{\bs{Q}}\bs{R}^{H}+\tilde{\bs{W}}_1,
\end{equation}
where the matrix $\tilde{\bs{W}}_1$ represents AWGN during Phase 1. Letting $\bs{H}\triangleq[\bar{\bs{h}}\ \ \check{\bs{h}}]$, $\bs{Q}\triangleq[\bar{\bs{Q}}\ \ \check{\bs{Q}}]$, $\bs{S}_1\triangleq[\bs{s}_1\ \ \bs{s}_1^{*}]$ and $\tilde{\bs{R}}\triangleq[\bs{R}\ \ \bs{R}^{*}]$, we rewrite~\eqref{EqY0} as
\begin{equation}
\label{EqY0}
\bs{Y}_1=\bs{H}\bs{S}_1^{T}+\bs{Q}\tilde{\bs{R}}^{T}+\tilde{\bs{W}}_1.
\end{equation}

To represent the signal in vector form, we let $\bs{y}_1\triangleq \mbox{Vec}(\bs{Y}_1)$, $\bs{h}\triangleq \mbox{Vec}(\bs{H})$, $\bs{q}\triangleq \mbox{Vec}(\bs{Q})$ and $\tilde{\bs{w}}_1\triangleq \mbox{Vec}(\bs{W}_1)$. We also define the matrix $\bs{P}\triangleq[\bs{S}\ \ \tilde{\bs{R}}]$ and the matrix $\bs{A}_1\triangleq\bs{P}\otimes\bs{I}_M$. The desired channel parameters are collected into the vector $\bs{\rho}\triangleq[\bs{h}^{T},\bs{q}^{T}]^{T}$.  
It can be verified that
\begin{equation}
\bs{y}_1=\bs{A}_1\bs{\rho}+\tilde{\bs{w}}_1.
\end{equation}
Hence, we can estimate the vector $\bs{\rho}$ using the LS criterion as 
$\hat{\bs{\rho}}=\bs{A}_1^{\dagger}\bs{y}_1$. We can extract from $\hat{\bs{\rho}}$ the estimates $\hat{\bar{\bs{h}}}$, $\hat{\check{\bs{h}}}$, $\hat{\bar{\bs{Q}}}$, $\hat{\check{\bs{Q}}}$.

Following Theorem 2 in~\cite{10271380}, the MSE of the above estimator is minimized if $\bs{P}^{H}\bs{P}$ is a diagonal matrix. Hence, optimal pilot design requires $\bs{s}_1$, $\bs{s}_1^{*}$, $\bs{r}_1,\bs{r}_1^{*},\hdots,\bs{r}_M,\bs{r}_M^{*}$ to be all mutually orthogonal. The requirement for the $M+1$ pilot sequences to be mutually orthogonal as well as orthogonal to their own complex conjugates and the complex conjugates of the other sequences guides the selection towards DFT sequences, which are capable of satisfying these requirements exactly, without requiring large sequence lengths. Other types of complex sequences, like ZC sequences, violate some of the orthogonality requirements (e.g., orthogonality between the sequence and its complex conjugate), which would result in non-resolvable error floors. Hence, the pilot sequences are obtained by selecting different columns of the DFT matrix. We let $\bs{F}$ be the $N_1\times N_1$ DFT matrix, whose $(i,j)th$ element is $1/\sqrt{N_1}e^{-\jmath2\pi (i-1)(j-1)/N_1}$. Requiring orthogonality between the sequences and their conjugates affects the selection of the DFT size. For each column of the DFT matrix, its complex conjugate is also a column of the DFT matrix. Hence, after excluding the real-valued columns, we have $(N_1-2)/2$ columns. Since we need to select $M+1$ columns from $\bs{F}$, this requires that $(N_1-2)/2\geq M+1$, i.e., $N_1\geq 2M+4$.

\subsubsection{Pilot Training Phase 2}

In this phase, we aim to estimate the channels $\bar{\bs{v}}_k$, $\check{\bs{v}}_k$, $\dot{\bs{v}}_k$ and $\ddot{\bs{v}}_k$ for $k=1,\hdots,K$. We assume that the AP is silent during this phase. Due to the coupling between mirror images transmitted by each tag, we divide this phase into two sub-phases. We transmit training sequences of length $N_2$ during each of the two sub-phases.
During the first sub-phase, the LU transmits the $N_2\times1$ vector $\bs{s}_2^{(1)}$, while the $k$th tag transmits the $N_2\times1$ vector $\bs{t}_k$. We let $\bs{T}_k\triangleq\mbox{Diag}(\bs{t}_k)$. The resulting $M\times N_1$ received signal matrix at the AP is 
\begin{equation}
\begin{split}
&\bs{Y}_2^{(1)}=\bar{\bs{h}}\bs{s}_2^{(1)T}+\check{\bs{h}}\bs{s}_2^{(1)H}+\sum\limits_{k=1}^{K}\Big(\bar{\bs{v}}_k\bs{s}_2^{(1)T}\bs{T}_k\\&+\check{\bs{v}}_k\bs{s}_2^{(1)H}\bs{T}_k^{*}+\dot{\bs{v}}_k\bs{s}_2^{(1)T}\bs{T}_k^{*}+\ddot{\bs{v}}_k\bs{s}_2^{(1)H}\bs{T}_k\Big)+\tilde{\bs{W}}^{(1)}_2,
\end{split}
\end{equation}
where $\tilde{\bs{W}}^{(1)}_2$ represents the AWGN matrix.

We set the LU signal as $\bs{s}_2^{(1)}\triangleq\bs{1}_{N_2}$. Hence, $\bs{s}_2^{(1)T}\bs{T}_k=\bs{1}_{N_2}\bs{T}_k=\bs{t}_k^{T}$. Furthermore, using the channel estimates from Phase 1, we cancel the direct signals from $\bs{Y}_2^{(1)}$, obtaining 
\begin{equation}
\begin{split}
&\tilde{\bs{Y}}_2^{(1)}\triangleq\bs{Y}_2^{(1)}-\hat{\bar{\bs{h}}}\bs{s}_2^{(1)T}-\hat{\check{\bs{h}}}\bs{s}_2^{(1)H}\\&
=\sum\limits_{k=1}^{K}\left(\bar{\bs{v}}_k+\ddot{\bs{v}}_k\right)\bs{t}_k^{(1)T}+\left(\check{\bs{v}}_k+\dot{\bs{v}}_k\right)\bs{t}_k^{(1)H}+\bs{W}^{(1)}_2+\bs{E}^{(1)}_2,
\end{split}
\end{equation}
where $\bs{E}^{(1)}_2$ is the residual error from the estimation of $\bar{\bs{h}}$ and $\check{\bs{h}}$. This makes it possible to estimate the aggregate vectors $\tilde{\bar{\bs{v}}}_k\triangleq\bar{\bs{v}}_k+\ddot{\bs{v}}_k$ and $\tilde{\check{\bs{v}}}_k\triangleq\check{\bs{v}}_k+\dot{\bs{v}}_k$.
We let
\begin{equation}
\bs{T}\triangleq\begin{bmatrix}
\bs{t}_1&\hdots&\bs{t}_K&\bs{t}_1^{*}&\hdots&\bs{t}_K^{*}
\end{bmatrix},
\end{equation}
and $\bs{A}_2\triangleq\bs{T}\otimes\bs{I}_M$. We also define the vector $\tilde{\bs{v}}^{(1)}\triangleq\big[\tilde{\bar{\bs{v}}}_1^{T},\hdots,\tilde{\bar{\bs{v}}}_K^{T},\tilde{\check{\bs{v}}}_1^{T},\hdots,\tilde{\check{\bs{v}}}_K^{T}\big]^{T}$. Hence, the vector $\tilde{\bs{y}}_2^{(1)}\triangleq\mbox{Vec}\big(\tilde{\bs{Y}}_2^{(1)}\big)$ can be expressed as
\begin{equation}
\tilde{\bs{y}}_2^{(1)}=\bs{A}_2\tilde{\bs{v}}^{(1)}+\bs{w}^{(1)}_2+\bs{e}^{(1)}_2.
\end{equation}
where $\bs{w}^{(1)}_2\triangleq\mbox{Vec}(\bs{W}^{(1)}_2)$ and $\bs{e}^{(1)}_2\triangleq\mbox{Vec}(\bs{E}^{(1)}_2)$.
We thus estimate the vector $\tilde{\bs{v}}^{(1)}$ by 
$\hat{\tilde{\bs{v}}}^{(1)}=\bs{A}_2^{\dagger}\tilde{\bs{y}}_2^{(1)}$.

As in Phase 1, we need to ensure that $\bs{T}^{H}\bs{T}$ is a diagonal matrix. This can be achieved by selecting the vectors $\bs{t}_k$ from the columns of an $N_2$-point DFT matrix. To have a sufficient number of orthogonal columns, we need $N_2\geq 2K+2$.

In the second sub-phase, we use the same training sequences for the tags, but use LU signal $\bs{s}_2^{(2)}=\jmath\bs{1}_{N_2}$. After removing the direct channel interference, the received signal is 
\begin{equation}
\begin{split}
&\tilde{\bs{Y}}_2^{(2)}=\sum\limits_{k=1}^{K}\jmath\left(\bar{\bs{v}}_k-\ddot{\bs{v}}_k\right)\bs{t}_k^{T}+\jmath\left(\dot{\bs{v}}_k-\check{\bs{v}}_k\right)\bs{t}_k^{H}+\bs{W}^{(2)}_2+\bs{E}^{(2)}_2,
\end{split}
\end{equation} 
where $\tilde{\bs{W}}^{(2)}_2$ represents the AWGN matrix and $\bs{E}^{(2)}_2$ is the residual error from the estimation of $\bar{\bs{h}}$ and $\check{\bs{h}}$. It is now possible to estimate the vectors $\bar{\bar{\bs{v}}}_k\triangleq\bar{\bs{v}}_k-\ddot{\bs{v}}_k$ and $\bar{\check{\bs{v}}}_k\triangleq \dot{\bs{v}}_k-\check{\bs{v}}_k$.  
Defining the vector $\bar{\bs{v}}^{(2)}\triangleq\big[\bar{\bar{\bs{v}}}_1^{T},\hdots,\bar{\bar{\bs{v}}}_K^{T},\bar{\check{\bs{v}}}_1^{T},\hdots,\bar{\check{\bs{v}}}_K^{T}\big]^{T}$ and letting $\tilde{\bs{y}}_2^{(2)}\triangleq\mbox{Vec}(\tilde{\bs{Y}}_2^{(2)})$, we obtain 
\begin{equation}
\tilde{\bs{y}}_2^{(2)}=\jmath\bs{A}_2\bar{\bs{v}}^{(2)}+\bs{w}^{(2)}_2+\bs{e}^{(2)}_2.
\end{equation}
where $\bs{w}^{(2)}_2\triangleq\mbox{Vec}(\bs{W}^{(1)}_2)$ and $\bs{e}^{(2)}_2\triangleq\mbox{Vec}(\bs{E}^{(1)}_2)$. We may thus estimate the vector $\bar{\bs{v}}^{(2)}$ as
$\hat{\bar{\bs{v}}}^{(2)}=(\jmath\bs{A}_2)^{\dagger}\tilde{\bs{y}}_2^{(2)}$.

Once the vectors $\tilde{\bar{\bs{v}}}_k$, $\tilde{\check{\bs{v}}}_k$, $\bar{\bar{\bs{v}}}_k$ and $\bar{\check{\bs{v}}}_k$ have been estimated, the estimates for the individual vectors $\bar{\bs{v}}_k$,$\ddot{\bs{v}}_k$, $\check{\bs{v}}_k$ and $\dot{\bs{v}}_k$ can be readily obtained by noting that $\bar{\bs{v}}=\frac{1}{2}(\tilde{\bar{\bs{v}}}_k+\bar{\bar{\bs{v}}}_k)$, $\ddot{\bs{v}}_k=\frac{1}{2}(\tilde{\bar{\bs{v}}}_k-\bar{\bar{\bs{v}}}_k)$, $\dot{\bs{v}}_k=\frac{1}{2}(\tilde{\check{\bs{v}}}_k+\bar{\check{\bs{v}}}_k)$ and $\check{\bs{v}}_k=\frac{1}{2}(\tilde{\check{\bs{v}}}_k-\bar{\check{\bs{v}}}_k)$.

\subsubsection{Pilot Training Phase 3}

In this phase, we estimate $\bar{\bs{U}}_k$, $\check{\bs{U}}_k$, $\dot{\bs{U}}_k$ and $\ddot{\bs{U}}_k$ for $k=1,\hdots,K$, all of which are $M\times M$ matrices. We assume that the AP transmits while the LU remains silent. If the tags transmit together, it is extremely challenging to find training sequences that guarantee orthogonality among all the channels, since all the tags' signals are multiplied by the same AP's sequence. Using DFT sequences will not ensure orthogonality in this case because of the matrix channels. Therefore, we divide this phase into $K$ stages, during each of which only one tag is active while the others are silent. Furthermore, each stage is further divided in two sub-stages, during each of which the tag transmits a pilot sequence of length $N_3$.

During the first sub-stage of the $k$-th tag's transmission, the tag transmits the $N_3\times1$ training sequence $\bs{\varrho}_k$. The training sequences transmitted by the AP's $M$ antennas during the same interval are collected into the $N_3\times M$ matrix $\bs{R}_3\triangleq\big[\bs{r}_1^{(3)},\hdots,\bs{r}_M^{(3)}\big]$. We also let $\tilde{\bs{R}}_3\triangleq[\bs{R}_3\ \bs{R}_3^{*}]$. The corresponding received signal is

\begin{equation}
\begin{split}
\bs{Y}_3^{(k,1)}&=\bs{Q}\tilde{\bs{R}}_3^{T}+\Big(\bar{\bs{U}}_k\bs{R}_3^{T}\bs{\Lambda}_k+\check{\bs{U}}_k\bs{R}_3^{H}\bs{\Lambda}_k^{*}\\&+\dot{\bs{U}}_k\bs{R}_3^{T}\bs{\Lambda}_k^{*}+\ddot{\bs{U}}_k\bs{R}_3^{H}\bs{\Lambda}_k\Big)+\tilde{\bs{W}}_3^{(k,1)},
\end{split}
\end{equation}
where $\bs{\Lambda}_k\triangleq\mbox{Diag}(\bs{\varrho}_k)$ and $\tilde{\bs{W}}_3^{(k,1)}$ is the AWGN matrix. We cancel the self-interference terms from $\bs{Y}_3^{(k,1)}$, obtaining 
\begin{equation}
\begin{split}
&\tilde{\bs{Y}}_3^{(k,1)}\triangleq\bs{Y}_3^{(k,1)}-\hat{\bs{Q}}\tilde{\bs{R}}_3^{T}\\& =\Big(\bar{\bs{U}}_k\bs{R}_3^{T}\bs{\Lambda}_k+\check{\bs{U}}_k\bs{R}_3^{H}\bs{\Lambda}_k^{*}+\dot{\bs{U}}_k\bs{R}_3^{T}\bs{\Lambda}_k^{*}\\&+\ddot{\bs{U}}_k\bs{R}_3^{H}\bs{\Lambda}_k\Big)+\bs{W}_3^{(k,1)}+\bs{E}_3^{(k,1)},
\end{split}
\end{equation}
where $\bs{E}_3^{(k,1)}$ represents the residual error from the cancellation. Furthermore, we use as the tag's training sequence $\bs{\varrho}_k=\bs{1}_{L}$, resulting in $\bs{\Lambda}_k=\bs{I}_{N_3}$. Hence,
\begin{equation}
\begin{split}
\tilde{\bs{Y}}_3^{(k,1)}&=(\bar{\bs{U}}_k+\dot{\bs{U}}_k)\bs{R}_3^{T}+(\check{\bs{U}}_k+\ddot{\bs{U}}_k)\bs{R}_3^{H}+\bs{W}_3^{(k,1)}+\bs{E}_3^{(k,1)}.
\end{split}
\end{equation}

We let $\tilde{\bar{\bs{U}}}_k\triangleq \bar{\bs{U}}_k+\dot{\bs{U}}_k$, $\tilde{\check{\bs{U}}}_k\triangleq \check{\bs{U}}_k+\ddot{\bs{U}}_k$, $\tilde{\bar{\bs{u}}}_k\triangleq\mbox{Vec}(\tilde{\bar{\bs{U}}}_k)$, $\tilde{\check{\bs{u}}}_k\triangleq\mbox{Vec}(\tilde{\check{\bs{U}}}_k)$ and $\tilde{\bs{u}}_k\triangleq\big[\tilde{\bar{\bs{u}}}_k^{T},\tilde{\check{\bs{u}}}_k^{T}\big]^{T}$. We also let $\tilde{\bs{y}}_3^{(k,1)}\triangleq\mbox{Vec}\left(\tilde{\bs{Y}}_3^{(k,1)}\right)$, $\bs{w}_3^{(k,1)}\triangleq\mbox{Vec}\left(\bs{W}_3^{(k,1)}\right)$ and $\bs{e}_3^{(k,1)}\triangleq\mbox{Vec}\left(\bs{E}_3^{(k,1)}\right)$. Moreover, we let $\bs{A}_3\triangleq\tilde{\bs{R}}_3\otimes\bs{I}_M$. Therefore, we can express $\tilde{\bs{y}}_3^{(k,1)}$ as
\begin{equation}
\tilde{\bs{y}}_3^{(k,1)}=\bs{A}_3\tilde{\bs{u}}_k+\bs{w}_3^{(k,1)}+\bs{e}_3^{(k,1)}.
\end{equation}
Hence, we estimate $\tilde{\bs{u}}_k$ by
$\hat{\tilde{\bs{u}}}_k=\bs{A}_3^{\dagger}\tilde{\bs{y}}_3^{(k,1)}$.
To ensure that $\tilde{\bs{R}}_3^{H}\tilde{\bs{R}}_3$ is diagonal, we select the vectors $\bs{r}_1^{(3)},\hdots,\bs{r}^{(3)}_M$ as columns of the $N_3$-point DFT matrix, where $N_3\geq 2M+2$. 

In the second sub-stage, the $k$th tag's pilot sequence is set to $\jmath\bs{1}_{N_3}$. We also let $\bar{\bar{\bs{U}}}_k\triangleq \bar{\bs{U}}_k-\dot{\bs{U}}_k$ and $\bar{\check{\bs{U}}}_k\triangleq \ddot{\bs{U}}_k-\check{\bs{U}}_k$. The corresponding received after interference cancellation is 
\begin{equation}
\begin{split}
\tilde{\bs{Y}}_3^{(k,2)}&
=\jmath\bar{\bar{\bs{U}}}_k\bs{R}_3^{T}+\jmath\bar{\check{\bs{U}}}_k\bs{R}_3^{H}+\bs{W}_3^{(k,2)}+\bs{E}_3^{(k,2)},
\end{split}
\end{equation}
where $\tilde{\bs{W}}_3^{(k,2)}$ is the AWGN matrix and $\bs{E}_3^{(k,2)}$ is the residual error. We also let $\bar{\bar{\bs{u}}}_k\triangleq\mbox{Vec}(\bar{\bar{\bs{U}}}_k)$, $\bar{\check{\bs{u}}}_k\triangleq\mbox{Vec}(\bar{\check{\bs{U}}}_k)$ and $\bar{\bs{u}}_k\triangleq\big[\bar{\bar{\bs{u}}}_k^{T},\bar{\check{\bs{u}}}_k^{T}\big]^{T}$. Furthermore, letting $\tilde{\bs{y}}_3^{(k,2)}\triangleq\mbox{Vec}\left(\tilde{\bs{Y}}_3^{(k,2)}\right)$, $\bs{w}_3^{(k,2)}\triangleq\mbox{Vec}\left(\bs{W}_3^{(k,2)}\right)$ and $\bs{e}_3^{(k,2)}\triangleq\mbox{Vec}\left(\bs{E}_3^{(k,2)}\right)$, we have
\begin{equation}
\tilde{\bs{y}}_3^{(k,2)}=\jmath\bs{A}_3\bar{\bs{u}}_k+\bs{w}_3^{(k,2)}+\bs{e}_3^{(k,2)}.
\end{equation}
Hence, we estimate $\bar{\bs{u}}_k$ by $\hat{\bar{\bs{u}}}_k=(\jmath\bs{A}_3)^{\dagger}\tilde{\bs{y}}_3^{(k,1)}$. After estimating $\tilde{\bs{u}}_k$ and $\bar{\bs{u}}_k$, it is straightforward to obtain the estimates of the channels $\bar{\bs{U}}_k$, $\check{\bs{U}}_k$, $\dot{\bs{U}}_k$ and $\ddot{\bs{U}}_k$. 

\begin{figure}[t]

    \centering
    \captionsetup{type=figure, justification=centering}
    \includegraphics[width=0.7\linewidth]{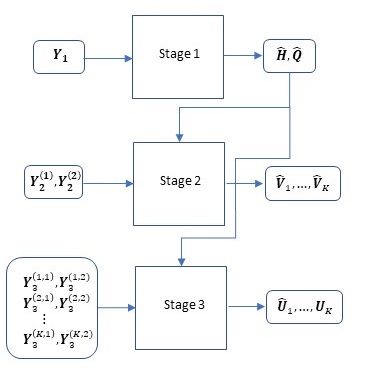}
    \caption{Flowchart of the proposed pilot-based estimator.}
    \label{flow_chart1}
\end{figure}

Fig.~\ref{flow_chart1} provides a flowchart showing the sequence and inputs/outputs of each stage of the pilot-based estimator. The total number of pilot transmission slots for all three phases is given by $N_1+2N_2+2KN_3$. Considering a typical channel coherence time of $100$ ms~\cite{9501035,7845587} and that AmBC systems have been shown to achieve data rates in the range of 1-5 Mbps~\cite{8642363,8368232}, the required number of pilots does not represent a significant overhead or bottleneck. For instance, considering $M=4$ antennas and $K=2$ tags, the number of required pilots is $N_{tot}=96$. Assuming a backscattering rate of $1$ Mpbs, the pilot transmission would represent less than $0.1\%$ of the coherence time.

\begin{figure}[t]
    \centering
    \captionsetup{type=figure, justification=centering}
    \includegraphics[width=0.8\linewidth]{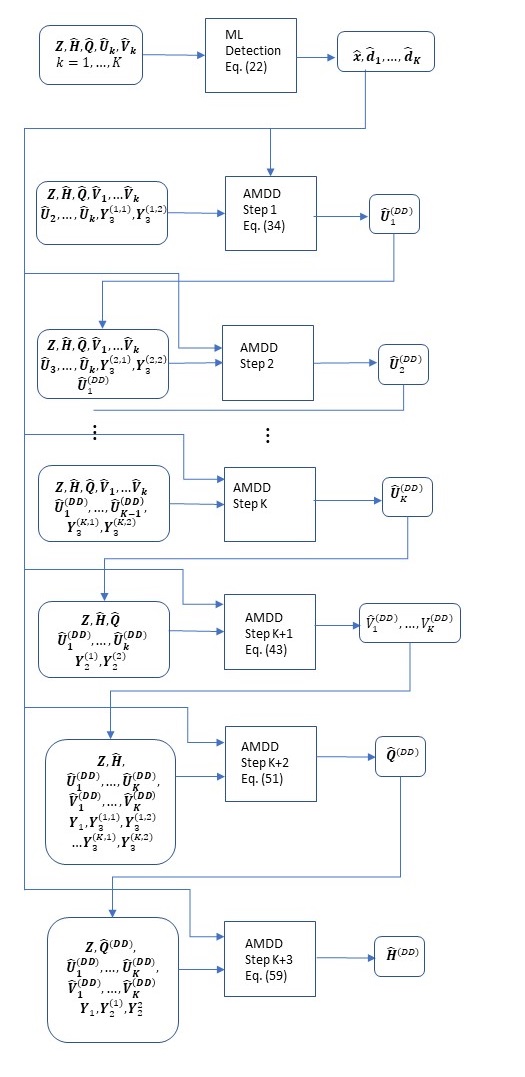}
    \caption{Flowchart of the proposed AMDD estimator.}
    \label{AMDD_diagram}
\end{figure}

\section{Semi-blind Alternate Maximization Decision-Directed Estimator}
\label{proposed_AMDD}

In this section, we propose a semi-blind estimator based on the DD criterion. Specifically, we use the pilot-based estimates acquired in Section~\ref{pilot_ML} to detect $D$ data transmissions. Another estimation stage is then applied whereby both the pilot sequences and the detected data symbols are used to re-estimate the channel coefficients. The detected data symbols are treated as if they are known pilots, enabling more accurate channel estimation, despite the possibility of error propagation from erroneous decisions. Due to the intractability of simultaneously estimating all the channel parameters, the re-estimation stage is performed by alternately maximizing the likelihood function with respect to the various channel vectors/matrices until all the parameters have been estimated. We refer to the proposed scheme as the alternate maximization DD (AMDD) estimator. 

The LU transmits $D$ data symbols, collected into the vector $\bs{x}\triangleq\big[x[1],\hdots,x[D]\big]^{T}$, where $\Exp{|x[i]|^2}=P_x$. The LU symbols are assumed to be generated from the set $\mathcal{S}$, of cardinality $\bar{\mathcal{S}}$. We also let $\bs{X}\triangleq[\bs{x}\ \  \bs{x}^{*}]$. In the same interval, each tag transmits $D$ symbols, collected into the vector $\bs{d}_k\triangleq \big[d_k[1],\hdots,d_k[D]\big]^{T}$. We let $\bs{D}_k\triangleq\mbox{Diag}(\bs{d}_k)$. The tag symbols are generated from the PSK set $\mathcal{D}$, of cardinality $\bar{\mathcal{D}}$. In contrast to the data symbols of the tags, those of the LU and the AP can be drawn from any constellation. Operating in full-duplex mode, the AP transmits the signals collected in the $D\times M$ matrix $\bs{C}\triangleq\big[\bs{c}[1],\hdots\bs{c}[D]\big]^{T}$, where $\bs{c}[n]$ is the $M\times1$ vector transmitted during the $n$th time slot. We let $\tilde{\bs{C}}\triangleq[\bs{C}\ \ \bs{C}^*]$. The received signal at the AP is 
\begin{equation}
\begin{split}
&\bs{Z}=\bs{H}\bs{X}^{T}+\bs{Q}\tilde{\bs{C}}^{T}+\sum\limits_{k=1}^{K}\Big(\bar{\bs{v}}_k\bs{x}^{T}\bs{D}_k+\check{\bs{v}}_k\bs{x}^{H}\bs{D}_k^{*}\\&+\dot{\bs{v}}_k\bs{x}^{T}\bs{D}_k^{*}+\ddot{\bs{v}}_k\bs{x}^{H}\bs{D}_k\Big)
+\sum\limits_{k=1}^{K}\Big(\bar{\bs{U}}_k\bs{C}^{T}\bs{D}_k+\check{\bs{U}}_k\bs{C}^{H}\bs{D}_k^{*}\\&+\dot{\bs{U}}_k\bs{C}^{T}\bs{D}_k^{*}+\ddot{\bs{U}}_k\bs{C}^{H}\bs{D}_k\Big)+\tilde{\bs{W}},
\end{split}
\end{equation}
where $\tilde{\bs{W}}$ represents the AWGN matrix. Before proceeding to the estimation stage, we need to detect the data symbols. Maximum-likelihood (ML) detection is discussed below.

\subsection{ML Detection}

In this step, we detect the vectors $\bs{x}$, $\bs{d}_1,\hdots,\bs{d}_K$. We begin by canceling the self-interference from $\bs{Z}$, using the pilot-based estimate $\hat{\bs{Q}}$ obtaining $\tilde{\bs{Z}}\triangleq\bs{Z}-\hat{\bs{Q}}\tilde{\bs{C}}^{T}$. Subsequently, ML data detection is given by
\begin{equation}
\label{MLdetection}
\begin{split}
&\{\hat{x}[\ell],\hat{d}_1[\ell],\hdots,\hat{d}_K[\ell]\}=\arg\min_{x\in\mathcal{S},d_1,\hdots,d_K\in\mathcal{D}}\Big\Vert\tilde{\bs{z}}[\ell]-\hat{\bar{\bs{h}}}x-\hat{\check{\bs{h}}}x^{*}\\
&-\sum\limits_{k=1}^{K}\Big(\hat{\bar{\bs{v}}}_kxd_k+\hat{\check{\bs{v}}}_kx^{*}d_k^{*}+\hat{\dot{\bs{v}}}_kxd_k^{*}+\hat{\ddot{\bs{v}}}_kx^{*}d_k\Big)\\
&-\sum\limits_{k=1}^{K}\Big(\hat{\bar{\bs{U}}}_k\bs{c}_k^{T}d_k+\hat{\check{\bs{U}}}_k\bs{c}_k^{H}d_k^{*}+\hat{\dot{\bs{U}}}_k\bs{c}_k^{T}d_k^{*}+\hat{\ddot{\bs{U}}}_k\bs{c}_k^{H}d_k\Big)\Big\Vert^2,
\end{split}
\end{equation}
for $\ell=1,\hdots,D$, where $\tilde{\bs{z}}[\ell]$ is the $\ell$th column of $\tilde{\bs{Z}}$.  The detected symbols are collected into the vectors $\hat{\bs{x}}$, $\hat{\bs{d}}_1,\hdots,\hat{\bs{d}}_K$.

\subsection{AMDD Estimation}

To simplify our notation, we let 
\begin{equation}
\bs{U}_k\triangleq
\begin{bmatrix}
\bar{\bs{U}}_k&\check{\bs{U}}_k&\ddot{\bs{U}}_k&\dot{\bs{U}}_k
\end{bmatrix},
\bs{V}_k\triangleq
\begin{bmatrix}
\bar{\bs{v}}_k&\check{\bs{v}}_k&\ddot{\bs{v}}_k&\dot{\bs{v}}_k
\end{bmatrix}.
\end{equation}
Furthermore, we let
\begin{equation}
\bar{\bs{D}}_k\triangleq\begin{bmatrix}
\bs{D}_k^{T}\bs{C}&\bs{D}_k^{H}\bs{C}^{*}&\bs{D}_k^{T}\bs{C}^{*}&\bs{D}_k^{H}\bs{C}
\end{bmatrix}^{T},
\end{equation}
and
\begin{equation}
\tilde{\bs{D}}_k\triangleq\begin{bmatrix}
\bs{D}_k^{T}\bs{x}&\bs{D}_k^{H}\bs{x}^{*}&\bs{D}_k^{T}\bs{x}^{*}&\bs{D}_k^{H}\bs{x}
\end{bmatrix}^{T}.
\end{equation}
We may thus express $\bs{Z}$ as
\begin{equation}
\begin{split}
&\bs{Z}=\bs{H}\bs{X}^{T}+\bs{Q}\tilde{\bs{C}}^{T}
+\sum\limits_{k=1}^{K}\bs{V}_k\tilde{\bs{D}}_k+\sum\limits_{k=1}^{K}\bs{U}_k\bar{\bs{D}}_k+\tilde{\bs{W}}.
\end{split}
\end{equation}

Estimation is performed via alternate maximization. We begin by re-estimating the matrices $\bs{U}_k$ for $k=1,\hdots,K$. We illustrate the estimation for $\bs{U}_1$. Prior to the estimation step, we cancel interference using the existing estimates, obtaining
\begin{equation}
\begin{split}
&\bs{Z}_u^{(1)}\triangleq\bs{Z}-\hat{\bs{H}}\hat{\bs{X}}^{T}-\hat{\bs{Q}}\tilde{\bs{C}}^{T}-\sum\limits_{k=1}^{K}\hat{\bs{V}}_k\hat{\tilde{\bs{D}}}_k-\sum\limits_{k=2}^{K}\hat{\bs{U}}_k\hat{\bar{\bs{D}}}_k\\
&=\bs{U}_1\bar{\bs{D}}_1+\tilde{\bs{W}}+\tilde{\bs{E}}_u^{(1)},
\end{split}
\end{equation}
where $\tilde{\bs{E}}_u^{(1)}$ is the residual error from the prior estimation and ML detection. We let $\bs{z}_u^{(1)}\triangleq\mbox{Vec}\left(\tilde{\bs{Z}}^{(1)}\right)$, $\bs{u}_1\triangleq\mbox{Vec}\left(\bs{U}_1\right)$, $\tilde{\bs{e}}_u^{(1)}\triangleq\mbox{Vec}\left(\tilde{\bs{E}}_u^{(1)}\right)$ and $\tilde{\bs{w}}\triangleq\mbox{Vec}(\tilde{\bs{W}})$. We also let $\bs{B}^{(k)}\triangleq\bar{\bs{D}}_k^{T}\otimes\bs{I}_M$. Hence, we obtain
\begin{equation}
\bs{z}_u^{(1)}=\bs{B}^{(1)}\bs{u}_1+\tilde{\bs{e}}_1+\tilde{\bs{w}}.
\end{equation}

In addition to $\bs{z}_u^{(1)}$, the pilot-carrying vectors $\tilde{\bs{y}}_3^{(1,1)}$ and $\tilde{\bs{y}}_3^{(1,2)}$ also bear information about $\bs{u}_1$. Letting
\begin{equation}
\bs{P}_3^{(1)}\triangleq\begin{bmatrix}
\bs{R}_3&\bs{R}_3^{*} &\bs{R}_3^{*}&\bs{R}_3
\end{bmatrix},
\end{equation}
\begin{equation}
\bs{P}_3^{(2)}\triangleq\begin{bmatrix}
\jmath\bs{R}_3&-\jmath\bs{R}_3^{*}&\jmath\bs{R}_3^{*} &-\jmath\bs{R}_3
\end{bmatrix},
\end{equation}
$\tilde{\bs{A}}_3^{(i)}\triangleq \bs{P}_3^{(i)}\otimes\bs{I}_M$, for $i=1,2$, we express $\tilde{\bs{y}}_3^{(1,i)}$ as
\begin{equation}
\tilde{\bs{y}}_3^{(1,i)}=\tilde{\bs{A}}_3^{(i)}\bs{u}_1+\bs{w}_3^{(1,i)}+\bs{e}_3^{(1,i)},
\end{equation}
for $i=1,2$. We now re-estimate $\bs{u}_1$ through the minimization
\begin{equation}
\begin{split}
\hat{\bs{u}}_1^{(DD)}&\triangleq\arg\min_{\bs{u}_1}\big\Vert\bs{z}_u^{(1)}-\hat{\bs{B}}^{(1)}\bs{u}_1\big\Vert^2+\sum_{i=1}^{2}\big\Vert\bs{y}_3^{(1,i)}-\tilde{\bs{A}}_3^{(i)}\bs{u}_1\big\Vert^2.
\end{split}
\end{equation}
The resulting AMDD estimator of $\bs{u}_1$ is given by
\begin{equation}
\label{AMDDu1}
\begin{split}
\hat{\bs{u}}_1^{(DD)}&=\Big(\hat{\bs{B}}^{(1)H}\hat{\bs{B}}^{(1)}+\sum_{i=1}^2\tilde{\bs{A}}_3^{(i)H}\tilde{\bs{A}}_3^{(i)}\Big)^{-1}\\ &\times\Big(\hat{\bs{B}}^{(1)H}\bs{z}_u^{(1)}+\sum_{i=1}^2\tilde{\bs{A}}_3^{(i)H}\bs{y}_3^{(1,i)}\Big).
\end{split}
\end{equation}

After estimating $\bs{u}_1$, the remaining vectors $\bs{u}_2,\hdots,\bs{u}_K$, are estimated in a similar fashion, using the most up-to-date estimates to cancel the interfering signal components.

We next estimate $\bs{v}_1,\hdots,\bs{v}_K$. We define the matrices
\begin{equation}
\bs{V}\triangleq\begin{bmatrix}
\bs{V}_1&\hdots&\bs{V}_K
\end{bmatrix},
\tilde{\bs{D}}\triangleq\begin{bmatrix}
\tilde{\bs{D}}_1^{T}&\hdots&\tilde{\bs{D}}_k^{T}
\end{bmatrix}^{T}.
\end{equation}

Removing the interference from $\bs{Z}$, we obtain
\begin{equation}
\begin{split}
&\bs{Z}_v\triangleq\bs{Z}-\hat{\bs{H}}\hat{\bs{X}}-\hat{\bs{Q}}\tilde{\bs{C}}-\sum\limits_{k=1}^{K}\hat{\bs{U}}_k^{(DD)}\hat{\bar{\bs{D}}}_k\\
&=\bs{V}\tilde{\bs{D}}+\tilde{\bs{W}}+\tilde{\bs{E}}_v,
\end{split}
\end{equation}
where $\tilde{\bs{E}}_v$ is the residual error. Letting $\bs{v}\triangleq\mbox{Vec}(\bs{V})$, $\bs{z}_v\triangleq\mbox{Vec}(\bs{Z}_v)$, $\bs{e}_v\triangleq\mbox{Vec}(\bs{E}_v)$ and $\bs{B}_v\triangleq \tilde{\bs{D}}^{T}\otimes\bs{I}_M$, we obtain
\begin{equation}
\bs{z}_v=\bs{B}_v\bs{v}+\tilde{\bs{e}}_v+\tilde{\bs{w}}.
\end{equation}

Moreover, since the vectors $\tilde{\bs{y}}_2^{(1)}$ and $\tilde{\bs{y}}_2^{(2)}$ also bear information about $\bs{v}$, they will be incorporated into the estimator of $\bs{v}$. We let $\tilde{\bs{T}}^{(1)}\triangleq [\tilde{\bs{T}}^{(1)}_1,\hdots\tilde{\bs{T}}^{(1)}_K]\otimes\bs{I}_M$, where
\begin{equation}
\tilde{\bs{T}}_k^{(1)}\triangleq\begin{bmatrix}
\bs{t}_k&\bs{t}_k^{*}&\bs{t}_k^{*}&\bs{t}_k
\end{bmatrix}.
\end{equation}
We also let $\tilde{\bs{T}}^{(2)}\triangleq [\tilde{\bs{T}}^{(2)}_1,\hdots\tilde{\bs{T}}_K^{(2)}]\otimes\bs{I}_M$, where
\begin{equation}
\tilde{\bs{T}}_k^{(2)}\triangleq\begin{bmatrix}
\jmath\bs{t}_k&-\jmath\bs{t}_k^{*}&\jmath\bs{t}_k^{*}&-\jmath\bs{t}_k
\end{bmatrix}.
\end{equation}
Hence, 
\begin{equation}
\tilde{\bs{y}}_2^{(i)}=\tilde{\bs{T}}^{(i)}\bs{v}+\bs{w}^{(i)}_2+\bs{e}^{(i)}_2,
\end{equation}
for $i=1,2$. The AMDD estimator of $\bs{v}$ is
\begin{equation}
\begin{split}
&\hat{\bs{v}}^{(DD)}=\arg\min_{\bs{v}}\big\Vert\bs{z}_v-\hat{\bs{B}}_v\bs{v}\big\Vert^2+\sum_{i=1}^{2}\big\Vert\bs{y}_2^{(i)}-\tilde{\bs{T}}^{(i)}\bs{v}\big\Vert^2.
\end{split}
\end{equation}
The resulting estimate of $\bs{v}$ is given by
\begin{equation}
\label{AMDDv}
\begin{split}
&\hat{\bs{v}}^{(DD)}\hs{0.5}=\hs{0.5}\Big(\hat{\bs{B}}_v^{H}\hat{\bs{B}}_v\hs{0.5}+\hs{0.5}\sum_{i=1}^2\tilde{\bs{T}}^{(i)H}\tilde{\bs{T}}^{(i)}\Big)^{\hs{0.5}-1}\hs{1}\Big(\hat{\bs{B}}_v^{H}\bs{z}_v\hs{0.5}+\hs{0.5}\sum_{i=1}^2\tilde{\bs{T}}^{(i)H}\bs{y}_2^{(i)}\Big).
\end{split}
\end{equation}

We next obtain the AMDD estimate of $\bs{Q}$. We let
\begin{equation}
\begin{split}
&\bs{Z}_q\triangleq\bs{Z}-\hat{\bs{H}}\hat{\bs{X}}-\hat{\bs{V}}^{(DD)}\hat{\tilde{\bs{D}}}-\sum\limits_{k=1}^{K}\hat{\bs{U}}^{(DD)}_k\hat{\bar{\bs{D}}}_k\\
&=\bs{Q}\tilde{\bs{C}}+\tilde{\bs{W}}+\tilde{\bs{E}}_q,
\end{split}
\end{equation}
where $\tilde{\bs{E}}_q$ is the residual error.
Letting $\bs{q}\triangleq\mbox{Vec}(\bs{Q})$, $\bs{z}_q\triangleq\mbox{Vec}(\bs{Z}_q)$, $\tilde{\bs{e}}_q\triangleq\mbox{Vec}(\tilde{\bs{E}}_q)$ and $\bs{B}_c\triangleq\tilde{\bs{C}}\otimes\bs{I}_M$, we obtain
\begin{equation}
\bs{z}_q=\bs{B}_c\bs{q}+\tilde{\bs{e}}_q+\tilde{\bs{w}}.
\end{equation}

We also incorporate $\bs{Y}_1$, $\bs{Y}_3^{(k,1)}$ and $\bs{Y}_3^{(k,2)}$ into the estimation of $\bs{Q}$. Removing the undesired signal components from $\bs{Y}_1$, we obtain
\begin{equation}
\begin{split}
\tilde{\bs{Y}}_1&\triangleq\bs{Y}_1-\hat{\bs{H}}\bs{S}_1^{T}=\bs{Q}\tilde{\bs{R}}^{T}+\tilde{\bs{E}}_1+\bs{W}_1.
\end{split}
\end{equation}
Furthermore, letting $\tilde{\bs{y}}_1\triangleq\mbox{Vec}(\tilde{\bs{Y}}_1)$, $\tilde{\bs{e}}_1\triangleq\mbox{Vec}(\tilde{\bs{E}}_1)$ and $
\bar{\bs{A}}_1\triangleq\tilde{\bs{R}}\otimes\bs{I}_M$,
we get
\begin{equation}
\tilde{\bs{y}}_1=\bar{\bs{A}}_1\bs{q}+\tilde{\bs{e}}_1+\bs{w}_1.
\end{equation}
We also let 
\begin{equation}
\begin{split}
\bar{\bs{Y}}_3^{(k,i)}&\triangleq\bs{Y}_3^{(k,i)}-\sum_{k=1}^{K}\hat{\bs{U}}_k^{(DD)}\bs{P}_3^{(i)}=\bs{Q}\bs{R}_3^{T}+\tilde{\bs{E}}_3^{(k,i)}+\bs{W}_3^{(k,i)},
\end{split}
\end{equation}
for $i=1,2$.
Moreover, letting $\bar{\bs{y}}_3^{(k,i)}\triangleq\mbox{Vec}\left(\bar{\bs{Y}}_3^{(k,i)}\right)$ and $\tilde{\bs{e}}_3^{(k,i)}\triangleq\mbox{Vec}\left(\tilde{\bs{E}}_3^{(k,i)}\right)$ for $i=1,2$, we obtain
\begin{equation}
\bar{\bs{y}}_3^{(k,i)}=\bs{A}_3\bs{q}+\tilde{\bs{e}}_3^{(k,i)}+\bs{w}_3^{(k,i)}.
\end{equation}

Therefore, we can estimate $\bs{q}$ using the minimization
\begin{equation}
\label{AMDDq}
\begin{split}
\hat{\bs{q}}^{(DD)}&=\arg\min_{\bs{q}}\big\Vert\bs{z}_q-\bs{B}_c\bs{q}\big\Vert^2+\big\Vert\tilde{\bs{y}}_1-\bar{\bs{A}}_1\bs{q}\big\Vert^2\\&
+\sum\limits_{k=1}^{K}\sum\limits_{i=1}^2\big\Vert\bar{\bs{y}}_3^{(k,i)}-\bs{A}_3\bs{q}\big\Vert^2.
\end{split}
\end{equation}
Hence,
\begin{equation}
\begin{split}
&\hat{\bs{q}}^{(DD)}=\left(\bs{B}_c^{H}\bs{B}_c+\bar{\bs{A}}_1^H\bar{\bs{A}}_1+2K\bs{A}_3^{H}\bs{A}_3\right)^{-1}\\&\times\Big(\bs{B}_c^{H}\bs{z}_q+\bar{\bs{A}}_1^{H}\tilde{\bs{y}}_1+\bs{A}_3^{H}\sum\limits_{k=1}^{K}\sum\limits_{i=1}^2\bar{\bs{y}}_3^{(k,i)}\Big).
\end{split}
\end{equation}

The last step of the AMDD estimator is to estimate $\bs{h}$. Removing the interference terms from $\bs{Z}$, we obtain 
\begin{equation}
\begin{split}
&\bs{Z}_h\triangleq\bs{Z}-\hat{\bs{Q}}^{(DD)}\tilde{\bs{C}}-\hat{\bs{V}}^{(DD)}\hat{\tilde{\bs{D}}}-\sum\limits_{k=1}^{K}\hat{\bs{U}}_k\hat{\bar{\bs{D}}}_k\\
&=\bs{H}\bs{X}^{T}+\tilde{\bs{W}}+\tilde{\bs{E}}_h,
\end{split}
\end{equation}
where $\tilde{\bs{E}}_h$ is the residual error.
Letting $\bs{z}_h\triangleq\mbox{vec}(\bs{Z}_h)$, $\bs{B}_h\triangleq\bs{X}\otimes\bs{I}_M$ and $\tilde{\bs{e}}_h\triangleq\mbox{Vec}(\tilde{\bs{E}}_h)$, we obtain
\begin{equation}
\bs{z}_h=\bs{B}_h\bs{h}+\tilde{\bs{e}}_h+\tilde{\bs{w}}.
\end{equation}

We also let,
\begin{equation}
\bar{\bs{Y}}_1\triangleq\bs{Y}_1-\hat{\bs{Q}}^{(DD)}\tilde{\bs{R}}^{T}=\bs{H}\bs{S}+\bar{\bs{E}}_1+\bs{W}_1,
\end{equation}
where $\bar{\bs{E}}_1$ is the residual error. Letting $\bar{\bs{S}}_1\triangleq\bs{S}_1\otimes\bs{I}_M$, $\bar{\bs{y}}_1\triangleq\mbox{Vec}(\bar{\bs{Y}}_1)$, $\bar{\bs{y}}_1\triangleq\mbox{Vec}(\bar{\bs{Y}}_1)$ and $\bar{\bs{e}}_1\triangleq\mbox{Vec}(\bar{\bs{e}}_1)$, we obtain
\begin{equation}
\bar{\bs{y}}_1=\bar{\bs{S}}_1\bs{h}+\bar{\bs{e}}_1+\bs{w}_1.
\end{equation}

We also remove the interference from the terms $\bs{Y}_2^{(1)}$ and $\bs{Y}_2^{(2)}$ using the latest estimates of $\bs{v}$. We let
\begin{equation}
\bar{\bs{Y}}_2^{(i)}\triangleq\bs{Y}_2^{(i)}-\sum_{k=1}^{K}\hat{\bs{V}}_k^{(DD)}\tilde{\bs{T}}_k^{(i)T}=\bs{H}\bs{S}_2^{(i)}+\bar{\bs{E}}_2^{(i)}+\bs{W}_2^{(i)},
\end{equation}
for $i=1,2$, where $\bs{S}_2^{(i)}\triangleq[\bs{s}_2^{(i)}\ \ \bs{s}_2^{(i)*}]$ and $\bar{\bs{E}}_2^{(i)}$ is the residual error. Letting $\bar{\bs{y}}_2^{(i)}\triangleq\mbox{Vec}(\bar{\bs{Y}}_2^{(i)})$, $\bar{\bs{e}}_2^{(i)}\triangleq\mbox{Vec}\big(\bar{\bs{E}}_2^{(i)}\big)$ and $\bar{\bs{S}}_2^{(i)}\triangleq\bs{S}_2^{(i)}\otimes\bs{I}_M$, we obtain
\begin{equation}
\bar{\bs{y}}_2^{(i)}=\bar{\bs{S}}_2^{(i)}\bs{h}+\bar{\bs{e}}_2^{(i)}+\bs{w}_2^{(i)}.
\end{equation}

Finally, we estimate $\bs{h}$ as
\begin{equation}
\begin{split}
\hat{\bs{h}}^{(DD)}&=\arg\min_{\bs{h}}\big\Vert\bs{z}_h-\hat{\bs{B}}_h\bs{h}\big\Vert^2+\big\Vert\bar{\bs{y}}_1-\bar{\bs{S}}_1\bs{h}\big\Vert^2\\&
+\sum_{i=1}^2\big\Vert\bar{\bs{y}}_2^{(i)}-\bar{\bs{S}}_2^{(i)}\bs{h}\big\Vert^2.
\end{split}
\end{equation}
Hence,
\begin{equation}
\label{AMDDh}
\begin{split}
&\hat{\bs{h}}^{(DD)}=\Big(\hat{\bs{B}}_h^{H}\hat{\bs{B}}_h+\bar{\bs{S}}_1^{H}\bar{\bs{S}}_1+\sum_{i=1}^2\bar{\bs{S}}_2^{(i)H}\bar{\bs{S}}_2^{(i)}\Big)^{-1}\\&\times\Big(\hat{\bs{B}}_h^{H}\bs{z}_h+\bar{\bs{S}}_1^{H}\bar{\bs{y}}_1+\sum_{i=1}^2\bar{\bs{S}}_2^{(i)H}\bar{\bs{y}}_2^{(i)}\Big).
\end{split}
\end{equation}

Fig.~\ref{AMDD_diagram} provides a flowchart showing the sequence and inputs/outputs of each stage of the AMDD estimator. The AMDD estimator provides superior accuracy to the pilot-based estimator, as demonstrated in Section~\ref{simulations}. However, the performance of the AMDD estimator is conditioned by the quality of the decisions made during the detection stage. This makes it prone to error propagation at low SNR. 

\section{Expectation Conditional Maximization Estimator}
\label{proposed_ECM_estimators}

In this section, we propose another semi-blind estimation scheme based on the ECM criterion. The ECM is an iterative approach that avoids hard detection of the data symbols. Instead, it evaluates the expectation of the complete-data log-likelihood function (LLF) using the posterior probabilities of the data symbols, computed in each iteration based on the currently available estimates. This constitutes the expectation step (E-step), which is followed by a maximization step (M-step), where estimates of the unknown parameters are updated.

The vectorized versions of all the channel parameters are collected into the $(2M+2M^2+4KM+4KM^2)\times1$ vector $\bs{\theta}\triangleq[\bs{h}^{T},\bs{q}^{T},\bs{v}^{T},\bs{u}^{T}]$. We also let $\bs{d}\triangleq[\bs{d}_1^{T},\hdots,\bs{d}_K^{T}]^{T}$. The complete data LLF of the received signal (if the data were perfectly known) is given by
\begin{equation}
\mathcal{L}(\bs{Y},\bs{Z}\ |\bs{x},\bs{d};\bs{\theta})=\mathcal{L}(\bs{Y};\bs{\theta})+\mathcal{L}(\bs{Z}\ |\bs{x},\bs{d};\bs{\theta}),
\end{equation}
where $\bs{Y}$ is the $M\times(N_1+2N_2+2KN_3)$ matrix aggregating all the received signals during pilot training, and the LLF for the pilot-carrying signals is given by
\begin{equation}
\begin{split}
&\mathcal{L}(\bs{Y};\bs{\theta})=-(N_1+2N_2+2KN_3)\log(\pi\tilde{\sigma}^2)\\&-\frac{1}{\tilde{\sigma}^2}\big\Vert\bs{Y}_1-\bs{H}\bs{S}_1^{T}-\bs{Q}\tilde{\bs{R}}^{T}\big\Vert^2\\&-\frac{1}{\tilde{\sigma}^2}\big\Vert\bs{Y}_2^{(1)}-\bs{H}\bar{\bs{S}}_2^T-\sum_{k=1}^K\bs{V}_k\tilde{\bs{T}}_k\big\Vert^2\\&-\frac{1}{\tilde{\sigma}^2}\big\Vert\bs{Y}_2^{(2)}-\jmath\bs{H}\bar{\bs{S}}_2^T-\sum_{k=1}^K\bs{V}_k\bar{\bs{T}}_k\big\Vert^2\\
&-\frac{1}{\tilde{\sigma}^2}\sum_{k=1}^{K}\Big(\big\Vert\bs{Y}_3^{(k,1)}-\bs{Q}\tilde{\bs{R}}_3^{T}-\sum_{k=1}^{K}\bs{U}_k\bs{P}_3^{(1)T}\big\Vert^2\\&+\big\Vert\bs{Y}_3^{(k,2)}-\bs{Q}\tilde{\bs{R}}_3^{T}-\sum_{k=1}^{K}\bs{U}_k\bs{P}_3^{(2)T}\big\Vert^2\Big).
\end{split}
\end{equation}
Moreover, the LLF for the data-carrying signals is given by
\begin{equation}
\begin{split}
&\mathcal{L}(\bs{Z} |\bs{x},\bs{d};\bs{\theta})=-D\log(\pi\tilde{\sigma}^2)-\frac{1}{\tilde{\sigma}^2}\sum\limits_{n=1}^{D}\big\Vert\bs{y}[n]-\bs{H}\tilde{\bs{x}}[n]^{T}\\&-\bs{Q}\bar{\bs{c}}[n]-\sum\limits_{k=1}^{K}\bs{V}_k\tilde{\bs{d}}_k[n]-\sum\limits_{k=1}^{K}\bs{U}_k\bs{\Upsilon}_k[n]\big\Vert^2,
\end{split}
\end{equation}
where $\tilde{\bs{x}}[n]\triangleq\big[x[n]\ x[n]^{*}\big]$, 
\begin{equation}
\tilde{\bs{d}}_k[n]\triangleq\begin{bmatrix}x[n]d_k[n]\\x[n]{*}d_k[n]^{*}\\x[n]^{*}d_k[n]\\x[n]d_k[n]^{*}\end{bmatrix},
\tilde{\bs{\Upsilon}}_k[n]\triangleq\begin{bmatrix}\bs{c}[n]d_k[n]\\ \bs{c}^{*}[n]d_k[n]^{*}\\ \bs{c}^{*}[n]d_k[n]\\ \bs{c}[n]d_k[n]^{*}\end{bmatrix}.
\end{equation}

The estimates of $\bs{V}_k$, $\bs{U}_k$, $\bs{Q}$, $\bs{H}$ during the $\ell$th ECM iteration are denoted by $\bs{V}_k^{(\ell)}$, $\bs{U}_k^{(\ell)}$, $\bs{Q}^{(\ell)}$ and $\bs{H}^{(\ell)}$, respectively. The vectorized estimates during the $\ell$th iteration are collected into the vector $\bs{\theta}^{(\ell)}$. The first step is to find the posterior probabilities of all the unknown data symbols. Letting $\bs{d}[n]\triangleq\big[d_1[n],\hdots,d_K[n]\big]^{T}$, the posterior probability for the $n$th data transmission during the $\ell$th iteration is  
\begin{equation}
B^{(\ell,n)}_{\lambda,\bs{\rho}}\triangleq P\left(x[n]=\lambda,\bs{d}[n]=\bs{\rho}|\bs{z}[n],\bs{\theta}^{(\ell)}\right),
\end{equation}
where $\lambda\in\mathcal{S}$ and $\bs{\rho}\in\mathcal{D}^K$, respectively.
We also let
\begin{equation}
\tilde{\bs{\rho}}_k\triangleq\begin{bmatrix}\xi\rho[k]&\xi^{*}\rho[k]^{*}&\xi^{*}\rho[k]&\xi\rho[k]^{*}\end{bmatrix}^{T},
\end{equation}
and
\begin{equation}
\tilde{\bs{\Gamma}}_k[n]\triangleq\begin{bmatrix}\bs{c}[n]\rho[k]\\ \bs{c}^{*}[n]\rho[k]^{*}\\ \bs{c}^{*}[n]\rho[k]\\ \bs{c}[n]\rho[k]^{*}\end{bmatrix}.
\end{equation}
Letting $\tilde{\bs{\lambda}}\triangleq[\lambda \ \ \lambda^{*}]$ and $\bar{\bs{c}}[n]\triangleq[\bs{c}[n]^T,\bs{c}[n]^{H}]^{T}$, we define
\begin{equation}
\begin{split}
G^{(\ell,n)}_{\lambda,\bs{\rho}}\triangleq\ &\mbox{exp}\Big(-\frac{1}{\tilde{\sigma}^2}\Big\Vert\bs{z}[n]-\bs{H}^{(\ell)}\tilde{\bs{\lambda}}^{T}-\bs{Q}^{(\ell)}\bar{\bs{c}}[n]\\&-\sum\limits_{k=1}^{K}\bs{V}_k\tilde{\bs{\rho}}_k-\sum\limits_{k=1}^{K}\bs{U}_k^{(\ell)}\tilde{\bs{\Gamma}}_k\Big\Vert^2\Big).
\end{split}
\end{equation}
Hence, the posterior probability is given by
\begin{equation}
\label{joint_posterior}
B^{(\ell,n)}_{\lambda,\bs{\rho}}=\frac{G^{(\ell,n)}_{\lambda,\bs{\rho}}}{\sum\limits_{\gamma\in\mathcal{S}}\sum\limits_{\bs{\eta}\in\mathcal{D}^{K}}G^{(\ell,n)}_{\gamma,\bs{\eta}}}.
\end{equation}
 We also let $\bs{\beta}\triangleq\begin{bmatrix}
\bs{\beta}_1^{T}& \hdots& \bs{\beta}_K^{T}\end{bmatrix}^{T}$ be the $KD\times1$ vector representing the possible data symbols of all the tags. Hence, we can define the posterior probability for the vector $\bs{z}$ as
\begin{equation}
 B^{(\ell)}_{\bs{\xi},\ \bs{\beta}}\triangleq P\left(\bs{x}=\bs{\xi},\bs{d}=\bs{\beta}\ \big|\bs{z},\bs{\theta}^{(\ell)}\right),
\end{equation}
where $\bs{\xi}$ is a $D\times1$ vector from $\mathcal{S}^{D}$.
\subsubsection{E-Step}In this step, we obtain the expectation of the complete-data LLF of the received signals using the computed posterior probabilities. Specifically, we obtain
\begin{equation}
\label{E_total}
\begin{split}
\mathcal{E}\left(\bs{\theta};\bs{\theta}^{(\ell)}\right)&\triangleq\Exp{\mathcal{L}(\bs{Y},\bs{Z}\ |\bs{x},\bs{d};\bs{\theta})\ |\bs{Y},\bs{Z};\bs{\theta}^{(\ell)}}\\
&=\mathcal{L}(\bs{Y};\bs{\theta})+\Sigma\left(\bs{\theta};\bs{\theta}^{(\ell)}\right),
\end{split}
\end{equation}
where 
\begin{equation}
\Sigma\left(\bs{\theta};\bs{\theta}^{(\ell)}\right)\triangleq\Exp{\mathcal{L}(\bs{Z}\ |\bs{x},\bs{d};\bs{\theta})\ |\bs{Z};\bs{\theta}^{(\ell)}}.
\end{equation}

We let $\tilde{\bs{\xi}}\triangleq[\bs{\xi}\ \ \bs{\xi}^{*}]$,
$\tilde{\bs{\beta}}_k\triangleq\begin{bmatrix}
\bs{\xi}\odot\bs{\beta}_k&\bs{\xi}^{*}\odot\bs{\beta}_k^{*}&\bs{\xi}^{*}\odot\bs{\beta}_k&\bs{\xi}\odot\bs{\beta}_k^{*}
\end{bmatrix}$, $\tilde{\bs{\beta}}\triangleq\begin{bmatrix}
\tilde{\bs{\beta}}_1&\hdots&\tilde{\bs{\beta}}_K\end{bmatrix}$, $\bs{\chi}_k\triangleq\mbox{Diag}(\bs{\beta}_k)$ and
$\tilde{\bs{\gamma}}_k\triangleq\begin{bmatrix}
\bs{\chi}_k^{T}\bs{C}&\bs{\chi}_k^{H}\bs{C}^{*}&\bs{\chi}_k^{T}\bs{C}^{*}&\bs{\chi}_k^{H}\bs{C}
\end{bmatrix}$.
It can be verified that
\begin{equation}
\begin{split}
\Sigma\left(\bs{\theta},\bs{\theta}^{\ell}\right)&=-D\log(\pi\tilde{\sigma}^2)-\frac{1}{\tilde{\sigma}^2}\sum\limits_{\bs{\xi}\in\mathcal{S}^{D}}\sum\limits_{\bs{\beta}\in\mathcal{D}^{KD}}B^{(\ell)}_{\bs{\xi},\bs{\beta}}\big\Vert\bs{z}\\&-\bs{G}_1\bs{h}-\bs{G}_2\bs{q}-\bs{G}_3\bs{v}-\sum\limits_{k=1}^{K}\bs{J}_k\bs{u}_k\big\Vert^2,
\end{split}
\end{equation}
where $\bs{G}_1=\tilde{\bs{\xi}}\otimes\bs{I}_M$, $\bs{G}_2\triangleq\tilde{\bs{C}}\otimes\bs{I}_M$, $\bs{G}_3\triangleq\tilde{\bs{\beta}}\otimes\bs{I}_M$ and $\bs{J}_k\triangleq\tilde{\bs{\gamma}}_k\otimes\bs{I}_M$.

We let $\bs{\Omega}_{ij}\triangleq\Exp{\bs{G}_i^{H}\bs{G}_j}$, for $i,j=1,\hdots,3$. We also let $\bs{\Omega}_{i}^{(k)}\triangleq\Exp{\bs{G}_i^{H}\bs{J}_k}$, for $i=1,2,3$. We also let $\bs{\Psi}_{i}\triangleq\Exp{\bs{G}_i}$, $\bs{\Pi}_k\triangleq\Exp{\bs{J}_k}$ and $\tilde{\bs{\Omega}}_{k,\ell}\triangleq\Exp{\bs{J}_k^{H}\bs{J}_{\ell}}$. Analytical expressions for the expectation terms can found using the posterior probability and corresponding marginal probabilities and are skipped due to space limitations. The result of the $E-Step$ is given by~\eqref{EM_expanded}.

\begin{figure*}[t]
\normalsize 
\begin{equation}
\label{EM_expanded}
\begin{split}
&\Sigma\left(\bs{\theta},\bs{\theta}^{\ell}\right)=-\frac{1}{\tilde{\sigma}^2}\Big(\bs{z}^{H}\bs{z}+\bs{h}^{H}\bs{\Omega}_{11}\bs{h}+\bs{q}^{H}\bs{\Omega}_{22}\bs{q}+\bs{v}^{H}\bs{\Omega}_{33}\bs{v}+\sum\limits_{k=1}^{K}\sum\limits_{\ell=1}^{K}\bs{u}_k^{H}\tilde{\bs{\Omega}}_{k\ell}\bs{u}_{\ell}-\bs{z}^{H}\bs{\Psi}_1\bs{h}-\bs{h}^{H}\bs{\Psi}_1^{H}\bs{z}-\bs{z}^{H}\bs{\Psi}_2\bs{q}-\bs{q}^{H}\bs{\Psi}_2^{H}\bs{z}\\&-\bs{z}^{H}\bs{\Psi}_3\bs{v}-\bs{v}^{H}\bs{\Psi}_3^{H}\bs{z}-\sum\limits_{k=1}^{K}\bs{z}^{H}\bs{\Pi}_{k}\bs{u}_k-\sum\limits_{k=1}^{K}\bs{u}_k^{H}\bs{\Pi}_{k}^{H}\bs{z}+\bs{h}^{H}\bs{\Omega}_{12}\bs{q}+\bs{q}^{H}\bs{\Omega}_{12}^{H}\bs{h}+\bs{h}^{H}\bs{\Omega}_{13}\bs{v}+\bs{v}^{H}\bs{\Omega}_{13}^{H}\bs{h}+\sum\limits_{k=1}^{K}\bs{h}^{H}\bs{\Omega}_{1}^{(k)}\bs{u}_k\\&+\sum\limits_{k=1}^{K}\bs{u}_k^{H}\bs{\Omega}_{1}^{(k)H}\bs{h}+\bs{q}^{H}\bs{\Omega}_{23}\bs{v}+\bs{v}^{H}\bs{\Omega}_{23}^{H}\bs{q}+\sum\limits_{k=1}^{K}\bs{q}^{H}\bs{\Omega}_{2}^{(k)}\bs{u}_k+\sum\limits_{k=1}^{K}\bs{u}_k^{H}\bs{\Omega}_{2}^{(k)H}\bs{q}+\sum\limits_{k=1}^{K}\bs{v}^{H}\bs{\Omega}_{3}^{(k)}\bs{u}_k+\sum\limits_{k=1}^{K}\bs{u}_k^{H}\bs{\Omega}_{3}^{(k)H}\bs{v}
\Big).
\end{split}
\end{equation}
\hrulefill
\end{figure*}

\subsubsection{Conditional Maximization}

When updating the estimate of a desired parameter (or group of parameters), the other parameters are fixed to their most up-to-date values, instead of being treated as unknown. The process is repeated alternately until all the parameters have been estimated. 

We start with the estimation of $\bs{u}_1$. We let $\bs{\theta}_{\bs{u}_1}^{(\ell)}$ be the version of the parameter vector $\bs{\theta}$ where all the parameters except $\bs{u}_1$ are set to their values from the $\ell$th iteration. 
To obtain the updated estimate of $\bs{u}_1^{(\ell+1)}$ of $\bs{u}_1$, we obtain the derivative of $\mathcal{E}\left(\bs{\theta}_{\bs{u}_1}^{(\ell)};\bs{\theta}^{(\ell)}\right)$ with respect to $\bs{u}_1^{*}$, where 
\begin{equation}
\frac{\partial\mathcal{E}\left(\bs{\theta}_{\bs{u}_1}^{(\ell)};\bs{\theta}^{(\ell)}\right)}{\partial\bs{u}_1^{*}}=\frac{\partial\mathcal{L}\left(\bs{Y};\bs{\theta}_{\bs{u}_1}^{(\ell)}\right)}{\partial\bs{u}_1^{*}}+\frac{\partial\Sigma\left(\bs{\theta}_{\bs{u}_1}^{(\ell)};\bs{\theta}^{(\ell)}\right)}{\partial\bs{u}_1^{*}}.
\end{equation}
Moreover,
\begin{equation}
\frac{\partial\mathcal{L}\left(\bs{Y};\bs{\theta}_{\bs{u}_1}^{(\ell)}\right)}{\partial\bs{u}_1^{*}}=\sum_{i=1}^2\Big(\tilde{\bs{A}}_{3}^{(i)H}\tilde{\bs{A}}_{3}^{(i)}\bs{u}_1-\tilde{\bs{A}}_{3}^{(i)H}\bs{y}_3^{(1,i)}\Big),
\end{equation}
and
\begin{equation}
\begin{split}
&\frac{\partial\Sigma\left(\bs{\theta}_{\bs{u}_1}^{(\ell)};\bs{\theta}^{(\ell)}\right)}{\partial\bs{u}_1^{*}}=\tilde{\bs{\Omega}}_{11}\bs{u}_1+\sum\limits_{\ell\neq\bs{1}}\tilde{\bs{\Omega}}_{1\ell}\bs{u}_{\ell}^{(\ell)}-\bs{\Pi}_{1}^{H}\bs{z}\\&+\bs{\Omega}_1^{(1)H}\bs{h}^{(\ell)}+\bs{\Omega}_{2}^{(1)H}\bs{q}^{(\ell)}+\bs{\Omega}_{3}^{(1)H}\bs{v}^{(\ell)}.
\end{split}
\end{equation}

Setting the derivative to zero and solving the resulting equation, we obtain
\begin{equation}
\label{u1_ECM}
\begin{split}
&\bs{u}_1^{(\ell+1)}=\Big(\sum_{i=1}^2\tilde{\bs{A}}_{3}^{(i)H}\tilde{\bs{A}}^{(i)}_{3}+\tilde{\bs{\Omega}}_{11}\Big)^{-1}\times\\&\Big(-\sum\limits_{k\neq\bs{1}}\tilde{\bs{\Omega}}_{1k}\bs{u}_{k}^{(\ell)}+\bs{\Pi}_{1}^{H}\bs{z}-\bs{\Omega}_1^{(1)H}\bs{h}^{(\ell)}-\bs{\Omega}_{2}^{(1)H}\bs{q}^{(\ell)}\\&-\bs{\Omega}_{3}^{(1)H}\bs{v}^{(\ell)}+\sum_{i=1}^2\tilde{\bs{A}}_{3}^{(i)H}\bs{y}_3^{(1,i)}\Big).
\end{split}
\end{equation}

A similar approach can be followed to find the ECM estimates of $\bs{u}_2,\hdots,\bs{u}_K$. This is followed by the estimation of $\bs{v}$. Obtaining the derivative of $\mathcal{E}\left(\bs{\theta}_{\bs{v}}^{(\ell)};\bs{\theta}^{(\ell)}\right)$ w.r.t $\bs{v}^*$ and setting it to zero, we obtain
\begin{equation}
\label{v_ECM}
\begin{split}
&\bs{v}^{(\ell+1)}=\left(\bs{\Omega}_{33}+\sum_{i=1}^2\tilde{\bs{T}}^{(i)H}\tilde{\bs{T}}^{(i)}\right)^{-1}\Big(\bs{\Psi}_3^{H}\bs{z}-\bs{\Omega}_{13}^{H}\bs{h}^{(\ell)}\\&-\bs{\Omega}_{23}^{H}\bs{q}^{(\ell)}-\sum_{k=1}^{K}\bs{\Omega}_3^{(k)}\bs{u}_k^{(\ell+1)}+\sum_{i=1}^2\tilde{\bs{T}}^{(i)H}\bs{y}_2^{(i)}\Big).
\end{split}
\end{equation}

We next obtain $\bs{q}^{(\ell+1)}$. Obtaining the derivative of $\mathcal{E}\left(\bs{\theta}_{\bs{q}}^{(\ell)};\bs{\theta}^{(\ell)}\right)$ w.r.t $\bs{q}^*$ and setting it to zero, we get
\begin{equation}
\label{q_ECM}
\begin{split}
&\bs{q}^{(\ell+1)}=\left(\bs{\Omega}_{22}+\bar{\bs{A}}_1^H\bar{\bs{A}}_1+2K\bs{A}_3^{H}\bs{A}_3\right)^{-1}\times\\&\Big(\bs{\Psi}_2^H\bs{z}-\bs{\Omega}_{12}^{H}\bs{h}^{(\ell)}-\bs{\Omega}_{23}\bs{v}^{(\ell+1)}+\sum_{k=1}^{K}\bs{\Omega}_2^{(k)}\bs{u}_k^{(\ell+1)}\\&+\bar{\bs{A}}_1^{H}\tilde{\bs{y}}_1+\bs{A}_3^{H}\sum\limits_{k=1}^{K}\Big(\bar{\bs{y}}_3^{(k,1)}+\bar{\bs{y}}_3^{(k,2)}\Big)\Big).
\end{split}
\end{equation}

Lastly, obtaining the derivative of $\mathcal{E}\left(\bs{\theta}_{\bs{h}}^{(\ell)};\bs{\theta}^{(\ell)}\right)$ w.r.t $\bs{h}^*$ and setting it to zero, we get
\begin{equation}
\label{h_ECM}
\begin{split}
&\bs{h}^{(\ell+1)}=\left(\bs{\Omega}_{11}+\bar{\bs{S}}_1^{H}\bar{\bs{S}}_1+\sum_{i=1}^2\bar{\bs{S}}_2^{(i)H}\bar{\bs{S}}_2^{(i)}\right)^{-1}\\&\times\Big(\bs{\Psi}_1^{H}\bs{z}-\bs{\Omega}_{12}\bs{q}^{(\ell+1)}-\bs{\Omega}_{13}\bs{v}^{(\ell+1)}\\&-\sum_{k=1}^{K}\bs{\Omega}_1^{(k)}\bs{u}_k^{(\ell+1)}+\bar{\bs{S}}_1^{H}\bar{\bs{y}}_1+\sum_{i=1}^2\bar{\bs{S}}_2^{(i)H}\bar{\bs{y}}_2^{(i)}\Big).
\end{split}
\end{equation}

The flowchart for the ECM estimator is generally similar to that of the AMDD estimator in Fig.~\ref{AMDD_diagram}, whereby the ML detection step is replaced with the E-step. It is excluded due to space limitations.

\section{Computational Complexity}
\label{computational_complexity}

In this section, we analyze the computational complexities of the three proposed estimators. The operations that can be performed off-line are excluded from the complexity calculation.

\subsection{Complexity of the Pilot-based Estimator}

The complexity of the pilot-based estimator depends on the three training phases. The complexity of phase 1 is dominated by the LS estimation of $\bs{\rho}$. However, the matrix $\bs{A}_1$ is entirely dependent on the fixed pilot sequences. Therefore, its pseudo-inverse needs to be computed only once and can be done off-line. Hence, the run-time complexity of Phase 1 is $O(M^3N_1)$. Phase 2 complexity is also dominated by the LS estimation of $\tilde{\bs{v}}^{(1)}$ the $\bar{\bs{v}}^{(2)}$,  which are $O(KM^2N_2)$. Phase 3 consists of $2K$ sub-phases. It is dominated by the complexity of the LS estimation of $\tilde{\bs{u}}_k$ and $\bar{\bs{u}}_k$, which are $O(M^3N_3)$. Since these operations are repeated $K$ times, the complexity will be $O(KM^3N_3)$. Combining all three phases, the total complexity of the pilot-based estimator is $O(M^3N_1+KM^2N_2+KM^3N_3)$. 

\subsection{Complexity of the AMDD Estimator}

The AMDD estimator involves a data detection stage prior to the estimation. The overall complexity of the ML data detection as outlined in~\eqref{MLdetection} is $O(D\bar{\mathcal{S}}K^{\bar{\mathcal{D}}+1}M^2)$. 

For the estimation part, the complexity of the estimation in~\eqref{AMDDu1} is $O(MD^2+M^6+M^3D+M^3N_3)$. Moreover, this step is repeated $K$ times resulting in an complexity of $O(K(MD^2+M^6+M^3D+M^3N_3))$. For the estimation of $\bs{v}$ in~\eqref{AMDDv}, the complexity is $O(KM^2D+K^3M^3+KM^2N_2)$. For the estimation of $\bs{q}$ in~\eqref{AMDDq}, the complexity is $O(M^6+M^3D+M^3N_1+KM^3N_3)$. As for the estimation of $\bs{h}$ in~\eqref{AMDDh}, the complexity is $O(M^3+M^2(D+N_1+N_2))$. Hence, the overall complexity of the AMDD estimator including both stages is $O(D\bar{\mathcal{S}}K^{\bar{\mathcal{D}}+1}M^2+KM^6+KM^3D+KM^3N_3+KMD^2+K^3M^3+KM^2N_2+M^3N_1)$.

\subsection{Complexity of the ECM Estimator}

The ECM involves three steps, 1) computing the posterior probabilities, 2) computing the expectation in the E-step and 3) performing the conditional maximization in the M-step.

It can be verified that the computation of the joint posterior probability in~\eqref{joint_posterior} is $O(M\bar{\mathcal{S}}+KM^2D\bar{\mathcal{D}}^K+D\bar{\mathcal{S}}\mathcal{D}^K)$. Moreover, the computation of the marginal probabilities needed for the E-step is $O(K^2D\bar{\mathcal{D}}^{K}\bar{\mathcal{S}})$.

The E-step involves obtaining the expectations in~\eqref{EM_expanded}. The complexity of the E-step is $O(K^2MD^2\bar{\mathcal{S}}\bar{\mathcal{D}}^2+K^2M^2D)$.

The last part of the ECM estimator is the M-step. Based on~\eqref{u1_ECM}, the estimation of $\bs{u}_k$, $k=1,\hdots,K$ is $O(KM^6+K^2M^4+KM^3D+KM^3N_3)$. Based on~\eqref{v_ECM}, the complexity for the estimation of $\bs{v}$ is $O(K^3M^3+KM^2D+KM^2N_2)$. Based on~\eqref{q_ECM}, the complexity of estimating $\bs{q}$ is $O(M^6+M^3D+M^4K+M^3N_1+M^3N_3+KMN_3)$. Based on~\eqref{h_ECM}, the complexity of estimating $\bs{h}$ is $O(KM^3+M^2D+M^2N_1+M^2N_2)$. Hence, the overall complexity of the M-step is $O(KM^6+K^3M^3+KM^3D+KM^3N_3+KM^2N_2+M^3N_1)$.

The overall complexity for an iteration of the ECM estimator including all three stages is thus $O(KM^2D\bar{\mathcal{D}}^K+K^2M^2D+K^2D\bar{\mathcal{D}}^K\bar{\mathcal{S}}+K^2MD^2\bar{\mathcal{S}}\bar{\mathcal{D}}^2+KM^6+K^3M^3+KM^3D+KM^3N_3+KM^2N_2+M^3N_1)$.

\section{Pilot-based Cramer-Rao Bound}
\label{Pilot_CRB}

In this section, we derive the pilot-based CRB. Letting $\bs{y}\triangleq\mbox{Vec}(\bs{Y})$, we can express $\bs{y}$ in terms of $\bs{\theta}$ as
\begin{equation}
\bs{y}=\bs{A}\bs{\theta}+\check{\bs{\omega}},
\end{equation}
where
\begin{equation}
\bs{A}\triangleq\begin{bmatrix}
\bs{S}_1&\tilde{\bs{R}}&\bs{0}&\bs{0}&\hdots&\hdots&\hdots&\bs{0}\\ \bs{S}_2&\bs{0}&\tilde{\bs{T}}^{(1)}&\bs{0}&\hdots&\hdots&\hdots&\bs{0}\\
\jmath\bs{S}_2&\bs{0}&\tilde{\bs{T}}^{(2)}&\bs{0}&\hdots&\hdots&\hdots&\bs{0}\\
\bs{0}&\bs{A}_3&\bs{0}&\tilde{\bs{A}}_3^{(1)}&\bs{0}&\hdots&\hdots&\bs{0}\\
\bs{0}&\bs{A}_3&\bs{0}&\tilde{\bs{A}}_3^{(2)}&\bs{0}&\hdots&\hdots&\bs{0}\\
\bs{0}&\bs{A}_3&\bs{0}&\bs{0}&\tilde{\bs{A}}_3^{(1)}&\bs{0}&\hdots&\bs{0}\\
\bs{0}&\bs{A}_3&\bs{0}&\bs{0}&\tilde{\bs{A}}_3^{(2)}&\bs{0}&\hdots&\bs{0}\\
\vdots&\vdots&\vdots&\vdots&\vdots&\ddots&\hdots&\bs{0}\\
\bs{0}&\bs{A}_3&\bs{0}&\bs{0}&\bs{0}&\bs{0}&\hdots&\tilde{\bs{A}}_3^{(1)}\\
\bs{0}&\bs{A}_3&\bs{0}&\bs{0}&\bs{0}&\bs{0}&\hdots&\tilde{\bs{A}}_3^{(2)}\\
\end{bmatrix},
\end{equation}
and $\check{\bs{w}}$ is the concatenated AWGN vector.

Letting $\bs{\Phi}\triangleq [\bs{\theta}^{T},\bs{\theta}^{H}]^{T}$, the FIM for our signal model is 
\begin{equation}
\begin{split}
\bs{I}_{\bs{\Phi}\bs{\Phi}}&\triangleq\mathbb{E}\left\{\frac{\partial \mathcal{L}(\bs{y};\bs{\theta})}{\partial\bs{\Phi}^{*}}\frac{\partial\mathcal{L}(\bs{y};\bs{\theta})^{H}}{\partial\bs{\Phi}^{*}}\right\}=\begin{bmatrix}\bs{I}_{\bs{\theta}\bs{\theta}} & \bs{I}_{\bs{\theta}\bs{\theta}^{*}}\\
 \bs{I}^{*}_{\bs{\theta}\bs{\theta}^{*}}&\bs{I}_{\bs{\theta}\bs{\theta}}^{*}
\end{bmatrix}.
\end{split}
\end{equation}
Furthermore, it can be verified that $\bs{I}_{\bs{\theta}\bs{\theta}}=\frac{1}{\tilde{\sigma}^2}\bs{A}^{H}\bs{A}$, while $\bs{I}_{\bs{\theta}\bs{\theta}^*}=\bs{0}$. Hence, $\bs{I}_{\bs{\Phi}\bs{\Phi}}=\frac{1}{\tilde{\sigma}^2}\begin{bmatrix}\bs{A}^{H}\bs{A} & \bs{0}\\
 \bs{0}&\bs{A}^{T}\bs{A}^{*}
\end{bmatrix}$.

We now define the real parameter vector $\bar{\bs{\theta}}\triangleq [\Re\{\bs{\theta}\}^{T},\Im\{\bs{\theta}\}^{T}]^{T}$. We also let $\bs{J}_{\bar{\bs{\theta}}\bar{\bs{\theta}}}$ be the corresponding FIM matrix for $\bar{\bs{\theta}}$. Then $\bs{J}_{\bar{\bs{\theta}}\bar{\bs{\theta}}}$ is related to $\bs{I}_{\bs{\Phi}\bs{\Phi}}$ by
$\bs{J}_{\bar{\bs{\theta}}\bar{\bs{\theta}}}=\mathcal{M}^{H}\bs{I}_{\bs{\Phi}\bs{\Phi}}\mathcal{M}$, where
$\mathcal{M}\triangleq\begin{bmatrix}\bs{I}_{\bar{L}}&\jmath\bs{I}_{\bar{L}}\\
\bs{I}_{\bar{L}}&-\jmath\bs{I}_{\bar{L}}
\end{bmatrix}$, and $\bar{L}\triangleq 2M+2M^2+4KM+4KM^2$. Therefore,
\begin{equation}
\bs{J}_{\bar{\bs{\theta}}\bar{\bs{\theta}}}=\frac{1}{\tilde{\sigma}^2}\begin{bmatrix}\bs{A}^{H}\bs{A}+\bs{A}^{T}\bs{A}^{*}&\jmath(\bs{A}^{H}\bs{A}-\bs{A}^{T}\bs{A}^{*})\\-\jmath{\bs{A}^{H}\bs{A}-\bs{A}^{T}\bs{A}^{*}}&\bs{A}^{H}\bs{A}+\bs{A}^{T}\bs{A}^{*}\end{bmatrix}.
\end{equation}
The pilot-based CRB is then 
$\mbox{CRB}_
{\bs{\theta}}=\mbox{tr}\left(\bs{J}_{\bar{\bs{\theta}}\bar{\bs{\theta}}}^{-1}\right)$.

\section{Semi-Blind Cramer-Rao Bound}
\label{SB_CRB}

Semi-blind estimation which incoporates both pilot and data symbols, requires the derivation of the semi-blind CRB. Since it is very tedious to obtain the exact semi-blind CRB, we consider the modified CRB (MCRB)~\cite{MCRB2000}, which is a looser bound. The MCRB first obtains the FIM conditioned on the data symbols, then applies the expectation with respect to the data symbols. Hence, the FIM in the MCRB is given by

\begin{equation}
\begin{split}
&\bs{J}_{\bs{\Phi}\bs{\Phi}}\triangleq\mathbb{E}_{\bs{Y},\bs{Z},\bs{x},\bs{d}}\left(\frac{\partial\mathcal{L}(\bs{Y},\bs{Z}|\bs{x},\bs{d};\bs{\theta})}{\bs{\Phi}^{*}}\frac{\partial\mathcal{L}(\bs{Y},\bs{Z}|\bs{x},\bs{d};\bs{\theta})^{H}}{\bs{\Phi}^{*}}\right)\\
&=\mathbb{E}_{\bs{x},\bs{d}}\mathbb{E}_{\bs{Y},\bs{Z}|\bs{x},\bs{d}}\left(\frac{\partial\mathcal{L}(\bs{Y},\bs{Z}|\bs{x},\bs{d};\bs{\theta})}{\bs{\Phi}^{*}}\frac{\partial\mathcal{L}(\bs{Y},\bs{Z}|\bs{x},\bs{d};\bs{\theta})^{H}}{\bs{\Phi}^{*}}\right)\hs{0.5}.
\end{split}
\end{equation}
It can be verified that $\bs{J}_{\bs{\Phi}\bs{\Phi}}=\begin{bmatrix}\bs{J}_{\bs{\theta}\bs{\theta}} & \bs{J}_{\bs{\theta}\bs{\theta}^{*}}\\
 \bs{J}^{*}_{\bs{\theta}\bs{\theta}^{*}}&\bs{J}_{\bs{\theta}\bs{\theta}}^{*}
\end{bmatrix}$. Moreover, $\bs{J}_{\bs{\theta}\bs{\theta}^{*}}=\bs{0}$, while 
$\bs{J}_{\bs{\theta}\bs{\theta}}=\bs{I}_{\bs{\theta}\bs{\theta}}+\tilde{\bs{J}}_{\bs{\theta}\bs{\theta}}$,
 where 
\begin{equation}
\tilde{\bs{J}}_{\bs{\theta}\bs{\theta}}=\frac{1}{\tilde{\sigma}^2}\begin{bmatrix}DP_x\bs{I}_{2M}&\bs{0}&\bs{0}&\bs{0}&\\ \bs{0}&\bs{B}_c^{H}\bs{B}_c&\bs{0}&\bs{0}&\\
\bs{0}&\bs{0}&DP_x\bs{I}_{4KM}&\bs{0}\\
\bs{0}&\bs{0}&\bs{0}&\tilde{\bs{B}} \end{bmatrix},
\end{equation}
where $\tilde{\bs{B}}$ is the $4KM^2\times4KM^2$ matrix given by
\begin{equation}
\tilde{\bs{B}}\triangleq\begin{bmatrix}
\check{\bs{C}}&\bs{0}&\hdots&\bs{0}\\
\bs{0}&\check{\bs{C}}&\bs{0}&\hdots\\
\vdots&\ddots&\ddots&\vdots\\
\bs{0}&\hdots&\hdots&\check{\bs{C}}
\end{bmatrix},
\end{equation}
where 
\begin{equation}
\check{\bs{C}}\triangleq\begin{bmatrix}
\bs{C}^{H}\bs{C}&\bs{0}&\bs{C}^{H}\bs{C}^{*}&\bs{0}\\
\bs{0}&\bs{C}^{T}\bs{C}^{*}&\bs{0}&\bs{C}^{T}\bs{C}\\
\bs{C}^{T}\bs{C}&\bs{0}&\bs{C}^{T}\bs{C}^{*}&\bs{0}\\
\bs{0}&\bs{C}^{H}\bs{C}^{*}&\bs{0}&\bs{C}^{H}\bs{C}
\end{bmatrix}.
\end{equation}
The semi-blind FIM for the real parameter vector $\bar{\bs{\theta}}$ is then 
$\bs{J}^{(SB)}_{\bar{\bs{\theta}}\bar{\bs{\theta}}}\triangleq\mathcal{M}^H\bs{J}_{\bs{\theta}\bs{\theta}}\mathcal{M}$, and  
$\mbox{CRB}_
{\bs{\theta}}^{(SB)}=\mbox{tr}\left(\left(\bs{J}^{(SB)}_{\bar{\bs{\theta}}\bar{\bs{\theta}}}\right)^{-1}\right)$.

\section{Simulation Results}
\label{simulations}

We use MATLAB to investigate the performance of the proposed estimators. The large-scale fading is modeled according to~\eqref{pathloss}, where $d_0=1$m, $d_R=\bar{d}_1=\bar{d}_2=30$m and $d_1=d_2=2$m. We use $f_c=915$ MHz, and $\gamma=2.5$. Small-scale fading is modeled as Nakagami-$m$ with $m=3$. The RSI coefficients are modeled using Rayleigh fading with $\sigma_i^2=-10$ dB. We set $\eta_k=0.6$ for $k=1,\hdots,K$. Pilot sequences are selected as columns of the DFT matrix, using $16$-pt and $32$-pt DFT. The data symbols of the LU and the tags are selected as QPSK symbols. The noise power is set to $-80$ dBm. For the I/Q imbalance, we assume no amplitude mismatch ($g_T=g_R=1$), while the phases $\phi_T$ and $\phi_R$ are generated randomly from the range $(0,1)$. Simulations are repeated over $100$ independent realizations of the channels and I/Q imbalance parameters. Unless specified otherwise, we use $K=2$ and $10$ ECM iterations.

\begin{figure} [t]
\centering
\includegraphics[width=3.8in, height=2.6in]{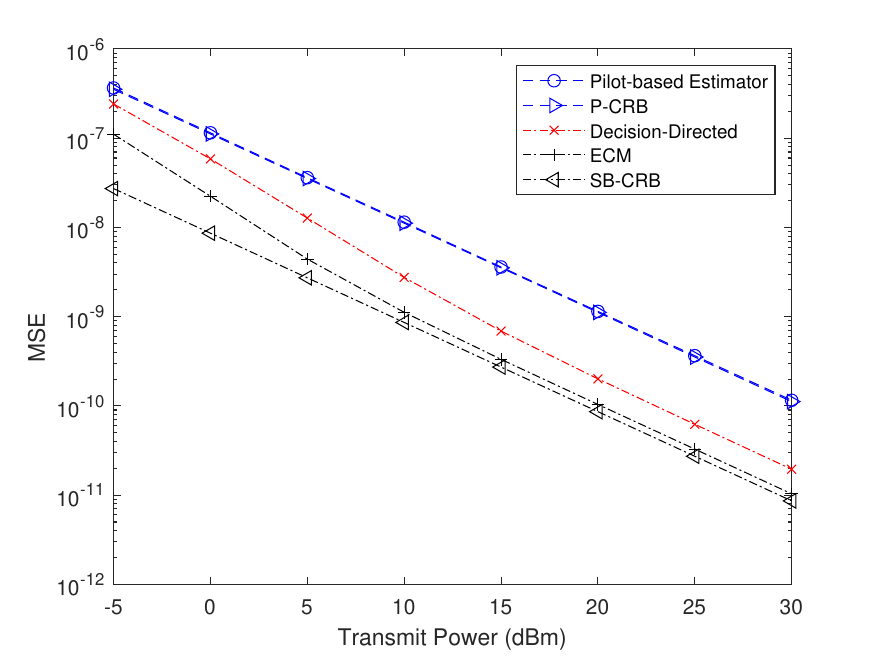}
\caption{MSE versus the transmit power ($M=4$, $N_1=N_2=N_3=16$, $D=200$).}
\label{MSEvsPT1}
\end{figure}

\begin{figure} [t]
\centering
\includegraphics[width=3.8in, height=2.6in]{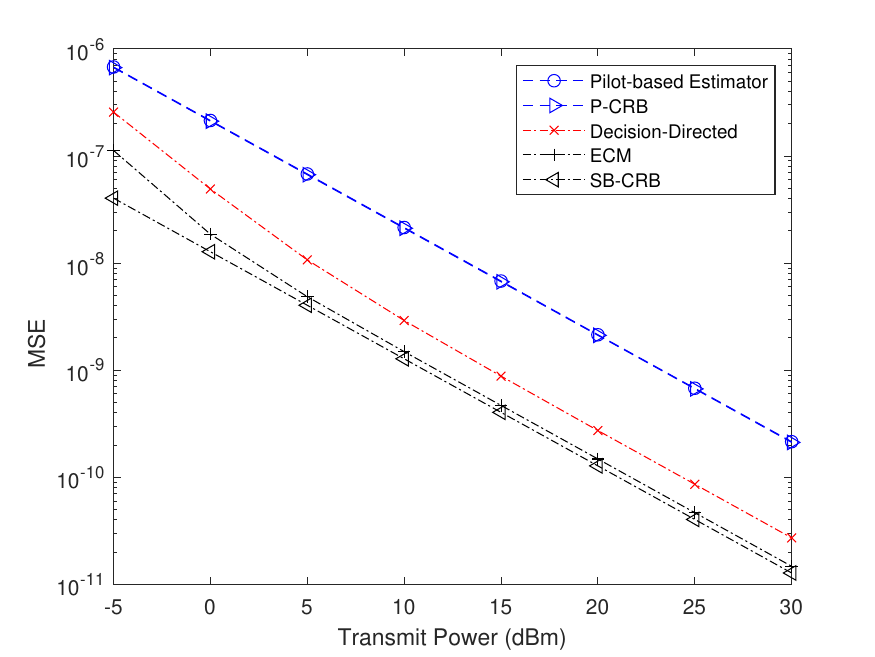}
\caption{MSE versus the transmit power ($M=8$, $N_1=N_2=N_3=32$, $D=500$).}
\label{MSEvsPT2}
\end{figure}

Fig.~\ref{MSEvsPT1} illustrates the MSE performance of the estimators versus the transmit power $P_T$ for $M=4$, pilot sizes $N_1=N_2=N_3 = 16$, and $D=200$. The proposed pilot-based estimator coincides with the P-CRB. Moreover, both proposed semi-blind estimators yield significantly higher estimation accuracy than the pilot-based estimator. For instance, at MSE of $10^{-9}$, the DD estimators provides a gain of 6.6 dB over the pilot-based estimator, while the ECM provides a gain of approximately $10.6$ dB. The ECM estimator performs very close to the SB-CRB starting at $P_T=10$ dBm, with a gap of just $0.5$ dB. On the other hand, the DD estimator still experiences a gap of $3$ dB from the SB-CRB even at high transmit power. Similar trends are observed in Fig.~\ref{MSEvsPT2}, where we vary the number antennas, length of pilot sequences and data size, using $M=8$, $N_1=N_2=N_3=32$ and $D=500$.

\begin{figure} [t]
\centering
\includegraphics[width=3.8in, height=2.6in]{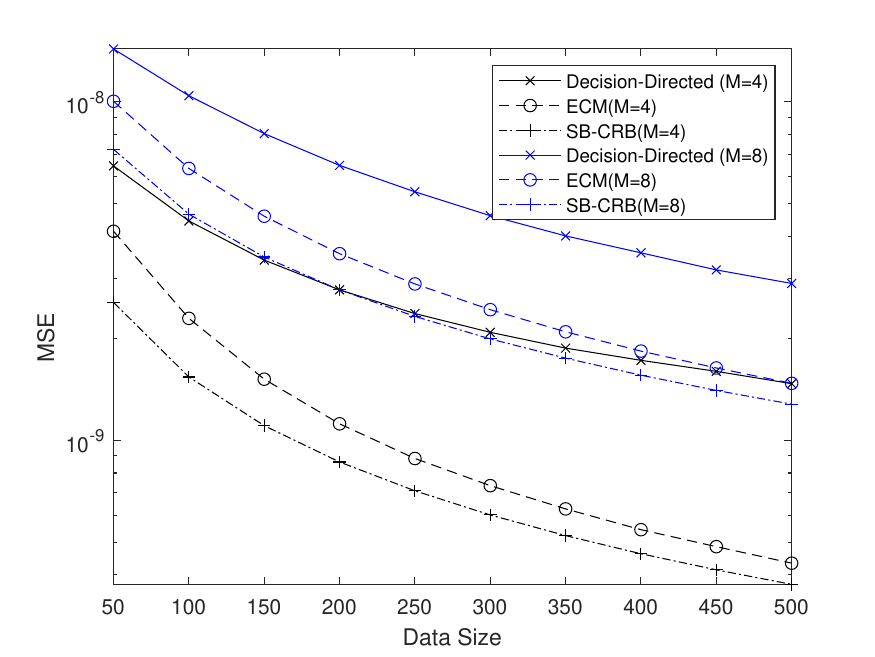}
\caption{MSE versus data size.}
\label{MSEvsData}
\end{figure}

Fig.~\ref{MSEvsData} depicts the MSE of the DD and ECM estimators versus the data size for $M=4$ and $M=8$ with $P_T=10$ dBm. We use $N_1=N_2=N_3=16$ for $M=4$ and $N_1=N_2=N_3=32$ for $M=8$. It is evident that performance of the semi-blind estimators improves with data size.

\begin{figure} [t]
\centering
\includegraphics[width=3.8in, height=2.6in]{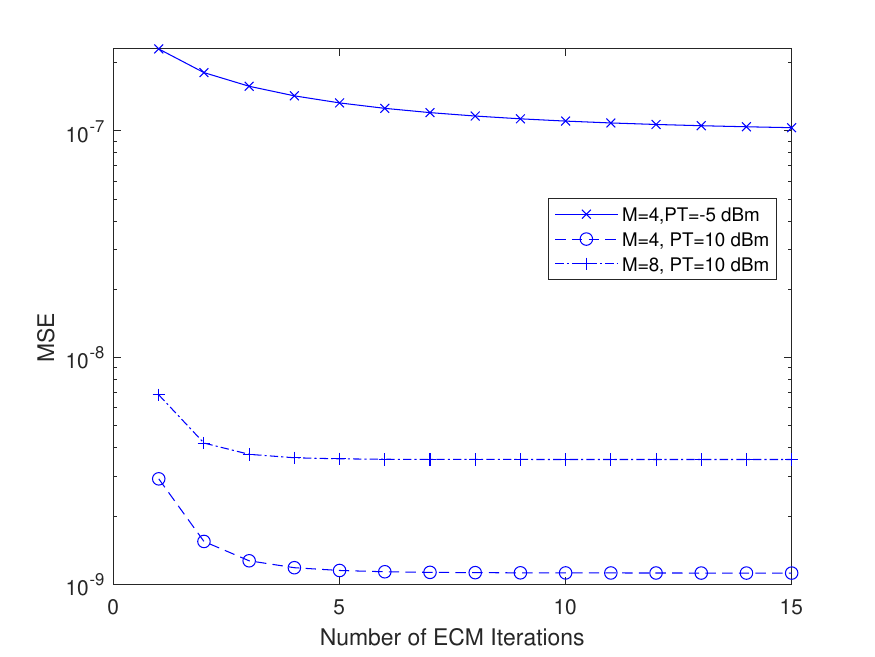}
\caption{MSE versus the number of ECM iterations.}
\label{MSEvsIterations}
\end{figure}

In Fig.~\ref{MSEvsIterations}, we investigate the performance of the ECM estimator versus the number of iterations. We consider three cases: ($M=4$, $P_T=-5$ dBm), ($M=4$, $P_T=10$ dBm) and  ($M=8$, $P_T=10$ dBm). For $P_T=10$ dBm, most of the improvement is achieved within the first $5$ iterations. On the other hand, for $P_T=-5$ dBm, most of the improvement is achieved within the first 10 iterations. The number of needed iterations is relatively small in both cases.

\begin{figure} [t]
\centering
\includegraphics[width=3.8in, height=2.6in]{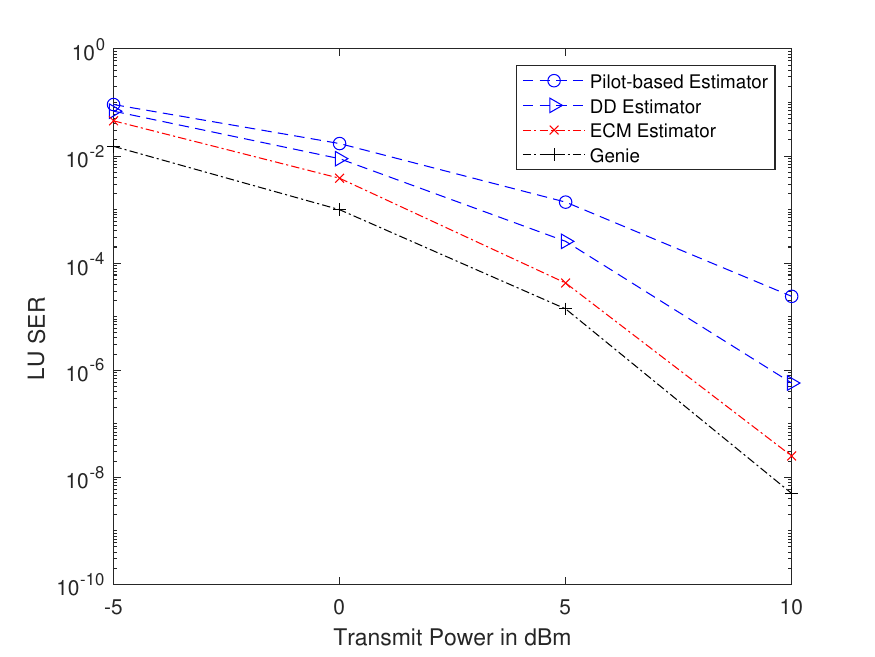}
\caption{LU SER of the three estimators and the ideal case.}
\label{SERs}
\end{figure}

In Fig.~\ref{SERs}, we plot the SER of the LU for the proposed estimators, using $M=4$, $N_1=N_2=N_3=16$ and $D=200$. As a benchmark, we show the performance under perfectly known channels (genie). The SER of all estimators is obtained using ML detection. It is observed again that both semi-blind estimators provide substantial performance improvements over the pilot-based estimator. At SER of $10^{-4}$, the DD estimator provides a gain of approximately $2.8$ dB over the pilot-based estimator, while the ECM provides a gain of approximately $4.4$ dB. The ECM performs closest to the ideal case, with a gap of just $1.4$ dB, followed by the DD with a gap of $3$ dB. 

\begin{figure} [t]
\centering
\includegraphics[width=3.8in, height=2.6in]{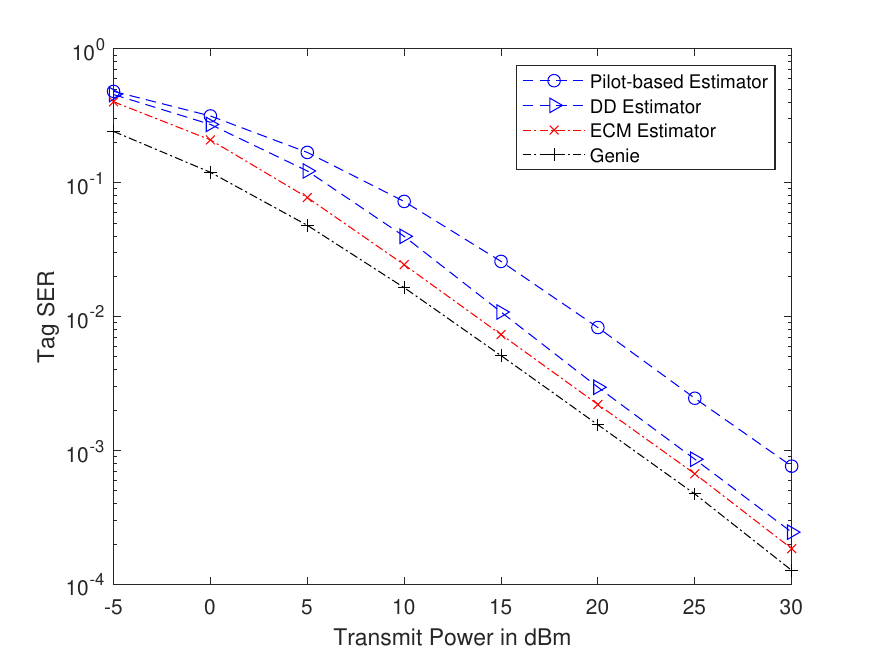}
\caption{Tag SER of the three estimators and the ideal case.}
\label{SERt}
\end{figure}

For the same setup, the SER of the tags is shown in Fig.~\ref{SERt}. Similar trends are observed. For SER of $10^{-3}$, the DD exhibits $5$ dB gain compared to the pilot-based estimator, while the ECM exhibits a gain of $6$ dB. The ECM is closest to the ideal performance with a gap of approximately $1.4$ dB.

\begin{figure} [htbp]
\centering
\includegraphics[width=3.8in, height=2.6in]{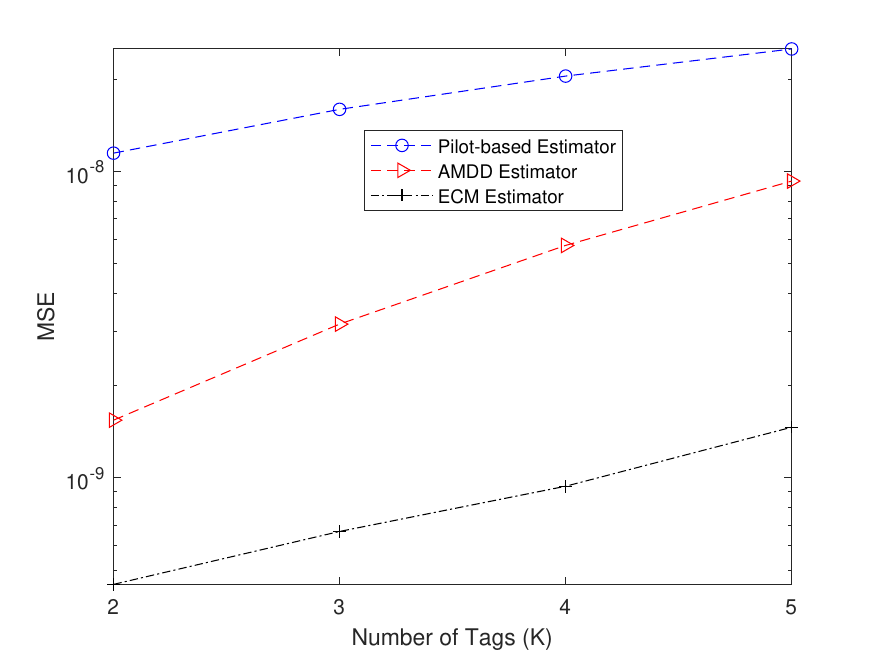}
\caption{MSE versus $K$ ($M=4$, $N=16$, $P_T=10$ dBm).}
\label{MSEvsK1}
\end{figure}

Fig.~\ref{MSEvsK1} shows the MSE of the estimators as number of tags $K$ increases from 2 to 5. The relative performance is maintained as $K$ increases, whereby the ECM provides the highest accuracy, whereas the AMDD estimator performs midway between the ECM and the pilot-based estimator. As expected, there is an increase in MSE as $K$ increases, which can be attributed both to the fact that there are more parameters to estimate, as well as because more tags means more inter-tag interference, resulting in higher MSE.

\section{Conclusions}
\label{conclusions}
In this paper, we considered the problem of joint estimation of the channel coefficients and I/Q imbalance for a multi-tag full-duplex AmBC system. The resulting estimation problem is highly challenging due to the large number of parameters, the coupling between them, the multiple mirror signal images and the presence of RSI. We proposed a training protocol and pilot-based estimator that guarantee orthogonality between the different signal components during pilot transmission, making it possible to estimate all cascaded channels without error floors. In addition, two semi-blind estimators were proposed, based on the AMDD strategy and the ECM framework. The pilot-based and semi-blind CRBs were obtained as performance benchmarks. Our simulations showed that the pilot-based estimator performance coincides with the corresponding CRB, while the ECM approaches its CRB. 

Both semi-blind estimators offered substantially higher estimation accuracy than the pilot-based estimator. An interesting tradeoff emerges, whereby the pilot-based estimator has the lowest complexity, the ECM has the best accuracy and highest complexity, and the AMDD offers an intermediate option. This offers valuable flexibility for the user to select the solution that best matches their needs and resources.


\bibliographystyle{IEEEtran}

\bibliography{IEEEabrv,biblio_fullduplex_updated}

\end{document}